\documentclass[%
 reprint,
superscriptaddress,
 amsmath,amssymb,
floatfix,
]{revtex4-2}

\usepackage[normalem]{ulem}
\usepackage{graphicx}
\usepackage{dcolumn}
\usepackage{bm}
\usepackage{comment}
\usepackage{physics}
\usepackage{amsmath}
\usepackage{lipsum}
\usepackage[dvipsnames]{xcolor}
\usepackage[T1]{fontenc}

\begin{document}

\title{Identification of  exciton complexes in a  charge-tuneable  Janus W$_\mathrm{Se}^\mathrm{S}$ monolayer}

\author{Matthew S. G. Feuer}
\thanks{These authors contributed equally to this work.}
\affiliation{Cavendish Laboratory, University of Cambridge, 19 J. J. Thomson Ave., Cambridge, CB3 0HE, UK}

\author{Alejandro R.-P. Montblanch}
\thanks{These authors contributed equally to this work.}
\affiliation{Cavendish Laboratory, University of Cambridge, 19 J. J. Thomson Ave., Cambridge, CB3 0HE, UK}

\author{Mohammed Sayyad}
\thanks{These authors contributed equally to this work.}
\affiliation{Materials Science and Engineering, School for Engineering of Matter, Transport and Energy, Arizona State University, Tempe, Arizona, 85287, USA}

\author{Carola M. Purser}
\affiliation{Cavendish Laboratory, University of Cambridge, 19 J. J. Thomson Ave., Cambridge, CB3 0HE, UK}
\affiliation{Cambridge Graphene Centre, University of Cambridge, 9 J. J. Thomson Ave., Cambridge, CB3 0FA, UK}

\author{Ying Qin}
\affiliation{Materials Science and Engineering, School for Engineering of Matter, Transport and Energy, Arizona State University, Tempe, Arizona, 85287, USA}

\author{Evgeny M. Alexeev}
\affiliation{Cambridge Graphene Centre, University of Cambridge, 9 J. J. Thomson Ave., Cambridge, CB3 0FA, UK}
\affiliation{Cavendish Laboratory, University of Cambridge, 19 J. J. Thomson Ave., Cambridge, CB3 0HE, UK}

\author{Alisson R. Cadore}
\affiliation{Cambridge Graphene Centre, University of Cambridge, 9 J. J. Thomson Ave., Cambridge, CB3 0FA, UK}

\author{Barbara L. T. Rosa}
\affiliation{Cambridge Graphene Centre, University of Cambridge, 9 J. J. Thomson Ave., Cambridge, CB3 0FA, UK}

\author{James Kerfoot}
\affiliation{Cambridge Graphene Centre, University of Cambridge, 9 J. J. Thomson Ave., Cambridge, CB3 0FA, UK}

\author{Elaheh Mostaani}
\affiliation{Cambridge Graphene Centre, University of Cambridge, 9 J. J. Thomson Ave., Cambridge, CB3 0FA, UK}

\author{Rados\l{}aw Kal\k{e}ba}
\affiliation{Cavendish Laboratory, University of Cambridge, 19 J. J. Thomson Ave., Cambridge, CB3 0HE, UK}

\author{Pranvera Kolari}
\affiliation{Materials Science and Engineering, School for Engineering of Matter, Transport and Energy, Arizona State University, Tempe, Arizona, 85287, USA}

\author{Jan Kopaczek}
\affiliation{Materials Science and Engineering, School for Engineering of Matter, Transport and Energy, Arizona State University, Tempe, Arizona, 85287, USA}

\author{Kenji Watanabe}
\affiliation{National Institute for Materials Science, Tsukuba, Ibaraki 305-0034, Japan}

\author{Takashi Taniguchi}
\affiliation{National Institute for Materials Science, Tsukuba, Ibaraki 305-0034, Japan}

\author{Andrea C. Ferrari}
\affiliation{Cambridge Graphene Centre, University of Cambridge, 9 J. J. Thomson Ave., Cambridge, CB3 0FA, UK}

\author{Dhiren M. Kara}
\affiliation{Cavendish Laboratory, University of Cambridge, 19 J. J. Thomson Ave., Cambridge, CB3 0HE, UK}

\author{Sefaattin Tongay}

\email{sefaattin.tongay@asu.edu}
\affiliation{Materials Science and Engineering, School for Engineering of Matter, Transport and Energy, Arizona State University, Tempe, Arizona, 85287, USA}

\author{Mete Atat\"{u}re}

\email{ma424@cam.ac.uk}
\affiliation{Cavendish Laboratory, University of Cambridge, 19 J. J. Thomson Ave., Cambridge, CB3 0HE, UK}

\begin{abstract}
Janus transition-metal dichalcogenide monolayers are fully artificial materials, where one plane of chalcogen atoms is replaced by chalcogen atoms of a different type. Theory predicts an in-built out-of-plane electric field,  giving rise to long-lived, dipolar excitons, while preserving direct-bandgap optical transitions in a uniform potential landscape. Previous Janus  studies had broad photoluminescence ($>$15~meV) spectra obfuscating their excitonic origin. Here, we identify the neutral, and negatively charged inter- and intravalley exciton transitions in Janus W$_\mathrm{Se}^\mathrm{S}$  monolayer with $\sim$6~meV optical linewidths. We combine a recently developed synthesis technique, with the integration of Janus monolayers into vertical heterostructures, allowing doping control. Further, magneto-optic measurements indicate that monolayer W$_\mathrm{Se}^\mathrm{S}$  has a direct bandgap at the K points. This work provides the foundation for applications such as nanoscale sensing, which relies on resolving excitonic energy shifts, and photo-voltaic energy harvesting, which requires efficient creation of long-lived excitons and integration into vertical heterostructures.
\end{abstract}

\maketitle

Layered materials are solids with strong intralayer bonds but only weak van der Waals coupling between layers~\cite{novoselov2005two}. These materials have a range of electronic~\cite{radisavljevic2011single}, optical~\cite{palacios2017large}, and topological~\cite{kou2017two} properties and can be combined in vertical heterostructures with pristine atomic interfaces despite mismatched lattice parameters~\cite{bonaccorso2012production, geim2013van, ferrari2015science, backes2020production}. Direct-bandgap semiconducting transition-metal dichalcogenide (TMD) monolayers are a class of layered material, which are particularly interesting due to their opto-electronic properties~\cite{mak2010atomically, xiao2012coupled, jones2013optical, koppens2014photodetectors}. Optical excitation creates excitons, i.e. bound electron-hole pairs, at the K and K' direct-bandgap edges~\cite{zeng2012valley,mak2012control}, while the strong spin-orbit interaction and broken inversion symmetry leads to coupling of the spin and valley degrees of freedom~\cite{wang2018colloquium}. Heterostructures comprising two different TMD monolayers  can have a type-II band alignment~\cite{kang2013band, gong2013band}, which localises electrons in one monolayer and holes in the other~\cite{mak2018opportunities}. This charge separation results in excitons with a permanent electric dipole moment~\cite{jauregui2019electrical} and long lifetime (up to 0.2~ms)~\cite{montblanch2021confinement}, due to a reduced overlap of electron and hole wavefunctions~\cite{jiang2021interlayer}. While such stacking configurations enable tuneability with layer angle and introduces emergent moir\'e physics~\cite{huang2022excitons}, they are also susceptible to an inhomogeneous potential landscape due to spatial variations in layer separation and twist angle~\cite{shabani2021deep,andrei2021marvels}.

Janus TMDs (J-TMD) are a new class of layered materials, first explored theoretically in 2013~\cite{cheng2013spin}, that promise  Rashba splitting~\cite{yao2017manipulation, hu2018intrinsic}; piezoelectric response~\cite{dong2017large,Cui2018}; and long-lived, dipolar excitons~\cite{Jin2018} in an intrinsically uniform potential landscape. To form a Janus monolayer, a conventional monolayer TMD, such as WSe$_2$, is altered to create W$_\mathrm{Se}^\mathrm{S}$ with Se atoms on one face and S atoms on the other, effectively placing a WSe$_2$/WS$_2$ interface within the monolayer. This artificially modified atomic ordering breaks the out-of-plane crystal symmetry and results in an in-built electric field~\cite{Li2017}, which separates the electron and hole wavefunctions and generates excitons with a permanent electric dipole~\cite{zheng2021excitonic}. Janus monolayers were first experimentally realised in 2017~\cite{zhang2017janus,lu2017janus}; however,  the excitonic origin of the optical transitions in J-TMDs has still not been clarified. One challenge is the broad 18-meV photoluminescence (PL) lineshape for the narrowest reported emission in J-TMDs, achieved via hexagonal boron nitride (hBN) encapsulation \cite{gan2022chemical}.  The second challenge is the absence, to-date, of integration of J-TMDs into electrically gated devices.

\begin{figure*}
\centering
\includegraphics[width=17.2cm]{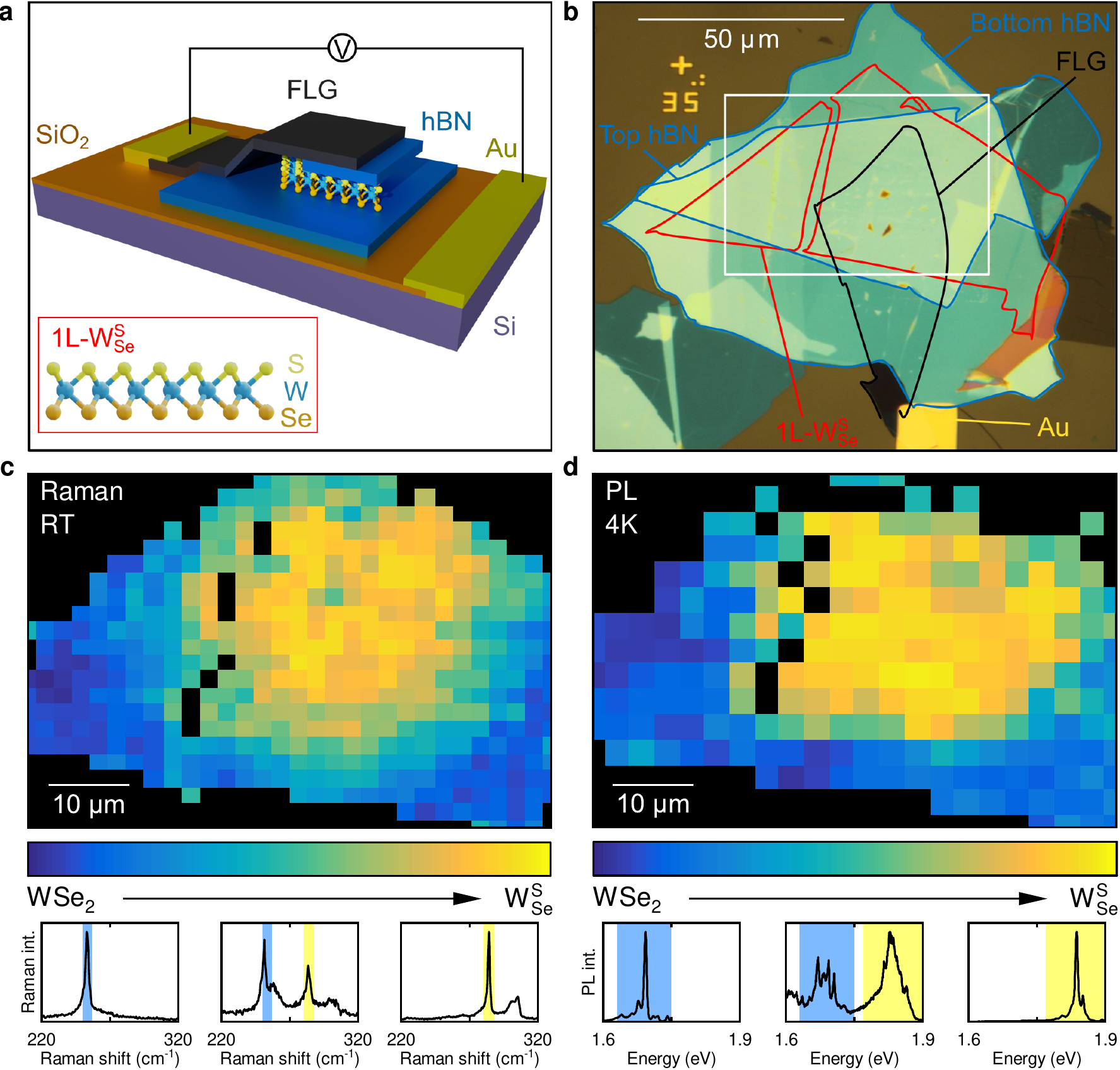}
\caption{Optical characterisation of the Janus 1L-W$_\mathrm{Se}^\mathrm{S}$ device.  \textbf{(a)} Illustration of the device. Janus 1L-W$_\mathrm{Se}^\mathrm{S}$ (inset) is encapsulated in ML-hBN (blue) and electrically contacted by FLG (black). The device is on a n$^{++}$ Si (purple)/SiO$_2$ (orange) substrate. Au contacts (yellow) allow a voltage to be applied between the FLG and Si. \textbf{(b)} Optical image of the device. 1L-W$_\mathrm{Se}^\mathrm{S}$ is outlined in red, the top and bottom ML-hBN in blue and FLG in black. \textbf{(c)} Raman map of the device, in the region highlighted by the white box in \textbf{b}, acquired at room temperature using 2.33~eV optical excitation. The colour coding shows the relative intensity between the 1L-WSe$_2$ $E^\prime$ + $A^\prime_1$ Raman mode (254~cm$^{-1}$), with 100\%  in blue,  and the Janus 1L-W$_\mathrm{Se}^\mathrm{S}$ $A^1_1$ Raman mode (284~cm$^{-1}$), with 100\%  in yellow. The substrate is shown in black. The arrow indicates conversion from 1L-WSe$_2$ to 1L-W$_\mathrm{Se}^\mathrm{S}$. Raman spectra from un-converted, partially converted and fully converted locations are shown below the Raman map, with the colour shading indicating the Raman modes above. \textbf{(d)} PL map of the device in the region highlighted by the white box in \textbf{b}, acquired at 4 K using 2.33~eV optical excitation. The colour coding shows the relative integrated PL emission intensity between the 1L-WSe$_2$ (1.63 to 1.75~eV), with 100\%  in blue, and Janus 1L-W$_\mathrm{Se}^\mathrm{S}$ (1.77 to 1.91~eV), with 100\% in yellow, spectral bands. The substrate is shown in black. The arrow indicates conversion from 1L-WSe$_2$ to 1L-W$_\mathrm{Se}^\mathrm{S}$. Representative normalised PL spectra from un-converted, partially converted and fully converted locations are shown below the PL map, with the colour shading indicating the spectral bands above.}
\label{fig:1}
\end{figure*}

In this work, we report the identification of  the neutral exciton (X$^0$), the negative inter- and intravalley trions (X$^-_\textrm{inter}$ and X$^-_\textrm{intra}$, respectively) and the X$^{-\prime}$ transition in a Janus W$_\mathrm{Se}^\mathrm{S}$ monolayer by using reflectance contrast  (RC) and PL spectroscopy. We confirm the Janus conversion of a monolayer exfoliated from flux-grown WSe$_2$ bulk crystal via Raman and PL spectroscopy over the flake. By  encapsulating monolayer W$_\mathrm{Se}^\mathrm{S}$ in hBN, we are able to measure the narrowest Janus emission (5.9 meV linewidth) reported. Charge-state control via direct contacting with graphene further enables identification of the negatively charged excitonic species. Furthermore, we measure the X$^0$ g factor to be $\sim$4, consistent with conventional TMDs. By establishing the excitonic origins of the spectral features in this material, we provide the basis for future work that will explore the novel properties of J-TMDs. 

\section*{Device characterisation}

Figure~\ref{fig:1}a is an illustration of one of our Janus devices. The silicon (Si) substrate is used as a back gate, separated from the Janus monolayer by SiO$_2$ and multilayer hBN (ML-hBN). A parent WSe$_2$ monolayer (1L-WSe$_2$) is exfoliated onto the ML-hBN. The 1L-WSe$_2$ is then converted into a Janus W$_\mathrm{Se}^\mathrm{S}$ monolayer  (1L-W$_\mathrm{Se}^\mathrm{S}$), with Se atoms on the bottom and S atoms on the top, by following a room-temperature in-\textit{situ} conversion technique (see Methods and Supplementary Notes S1, S2)~\cite{Tongay2020, Qin2022}. An additional ML-hBN transferred on top of the converted 1L-W$_\mathrm{Se}^\mathrm{S}$ encapsulates the flake, and a top gate comprised of few-layer graphene (FLG) electrically contacts the 1L-W$_\mathrm{Se}^\mathrm{S}$. Figure~\ref{fig:1}b shows an optical microscope image of the device, where the  1L-W$_\mathrm{Se}^\mathrm{S}$ is outlined in red, the bottom and top hBN in blue, and the FLG in black.

Figure~\ref{fig:1}c shows a Raman spectroscopy map of the device, acquired at room temperature using 2.33~eV optical excitation, in the region highlighted by the white box in Fig.~\ref{fig:1}b. The colour code indicates the relative intensity between the characteristic 1L-WSe$_2$ $E^\prime$ + $A^\prime_1$ Raman mode (blue) ~\cite{tonndorf2013photoluminescence} and the Janus 1L-W$_\mathrm{Se}^\mathrm{S}$ $A^1_1$ Raman mode (yellow)  ~\cite{petric2021raman}, with representative Raman spectra from regions with different degrees of Janus conversion shown below the Raman map (see S3). The Raman spectra from the large region ($\sim$~400~$\mu$m$^2$) of fully converted Janus 1L-W$_\mathrm{Se}^\mathrm{S}$ evidence that  the converted region is ordered Janus 1L-W$_\mathrm{Se}^\mathrm{S}$~\cite{petric2021raman} rather than a disordered alloy, which would show the representative Raman peaks of 1L-WSe$_2$ and 1L-WS$_2$~\cite{duan2016synthesis, Tongay2020}.

Figure~\ref{fig:1}d shows a PL map, acquired at 4~K using 2.33~eV optical excitation, in the same region of the device as in Fig.~\ref{fig:1}c. Similar to  Fig.~\ref{fig:1}c, the colour code shows the relative PL emission intensity between the distinct 1L-WSe$_2$ (blue) and Janus 1L-W$_\mathrm{Se}^\mathrm{S}$ (yellow) spectral bands~\cite{Tongay2020,Xinhui2021}. The PL map displays strong correlation with the Raman map in Fig.~\ref{fig:1}c, which validates our  assignment of the Janus 1L-W$_\mathrm{Se}^\mathrm{S}$ spectral band. Therefore, we focus on the exciton emission in the spatial region of full Janus conversion.

\section*{Identification of the neutral exciton}

\begin{figure}
\centering
\includegraphics[width=\linewidth]{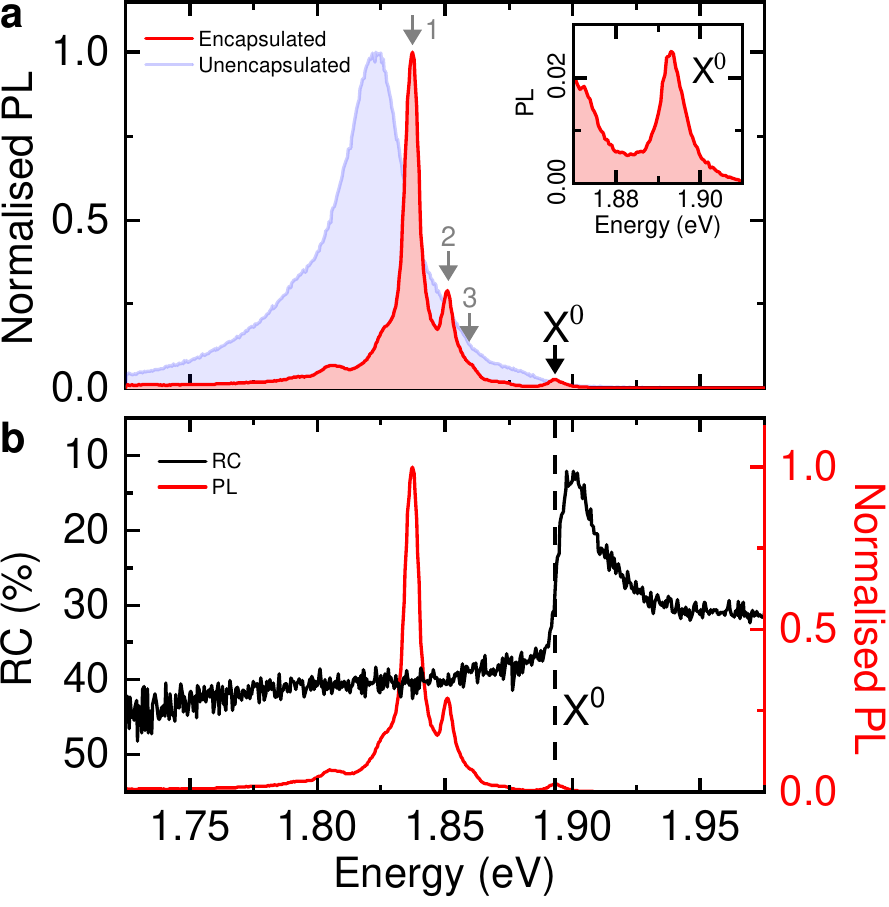}
\caption{Photoluminescence and reflectance contrast spectra of hBN-encapsulated 1L-W$_\mathrm{Se}^\mathrm{S}$. \textbf{(a)}  PL spectrum from the  encapsulated 1L-W$_\mathrm{Se}^\mathrm{S}$ device (red curve) compared to the PL spectrum from  unencapsulated 1L-W$_\mathrm{Se}^\mathrm{S}$ (blue curve). The spectra are normalised to the same peak height. The peaks labelled 1, 2 and 3 are present across the device. The inset shows the magnified PL spectrum around X$^0$. \textbf{(b)}~RC spectrum (black curve, left axis) from the  encapsulated 1L-W$_\mathrm{Se}^\mathrm{S}$ device compared to the PL spectrum at the same location (red curve, right axis). The black dashed line denotes the X$^0$ transition energy, 1.893 eV. All spectra were acquired at 4~K and the PL spectra under 2.33 eV
excitation.}
\label{fig:2}
\end{figure}

Encapsulation in hBN reduces the linewidths of PL peaks in conventional 1L-TMDs~\cite{Urbaszek2017,dean2010boron,man2016protecting}, thus allowing for the identification of excitonic species~\cite{barbone2018charge, li2018revealing}. Figure~\ref{fig:2}a compares a representative PL spectrum at 4~K from our ML-hBN encapsulated 1L-W$_\mathrm{Se}^\mathrm{S}$ device (red curve) to the spectrum from unencapsulated 1L-W$_\mathrm{Se}^\mathrm{S}$ on a Si/SiO$_2$ substrate (blue curve). The unencapsulated 1L-W$_\mathrm{Se}^\mathrm{S}$ has a broad spectrum, with a full width at half maximum (FWHM) on the order 30~meV, on par with the narrowest linewidth reported to-date for unencapsulated Janus TMDs~\cite{Qin2022}. In contrast, encapsulation with hBN allows us to resolve multiple spectral features with significantly reduced linewidths ($<$10~meV).

The  peaks  labelled 1, 2, 3 and X$^0$ are present in the 1L-W$_\mathrm{Se}^\mathrm{S}$ PL spectra across the whole device (see S3), indicating that these arise from intrinsic excitonic transitions. Since the highest-energy PL peak in both 1L-WSe$_2$ and 1L-WS$_2$  stems from neutral excitons~\cite{Urbaszek2017}, the  peak at 1.893~eV is a likely  candidate for the neutral exciton, X$^0$, in 1L-W$_\mathrm{Se}^\mathrm{S}$. To verify this, we directly probe excitonic absorption resonances using RC spectroscopy (see Methods)~\cite{mak2012control}.

\begin{figure}
\centering
\includegraphics[width=\linewidth]{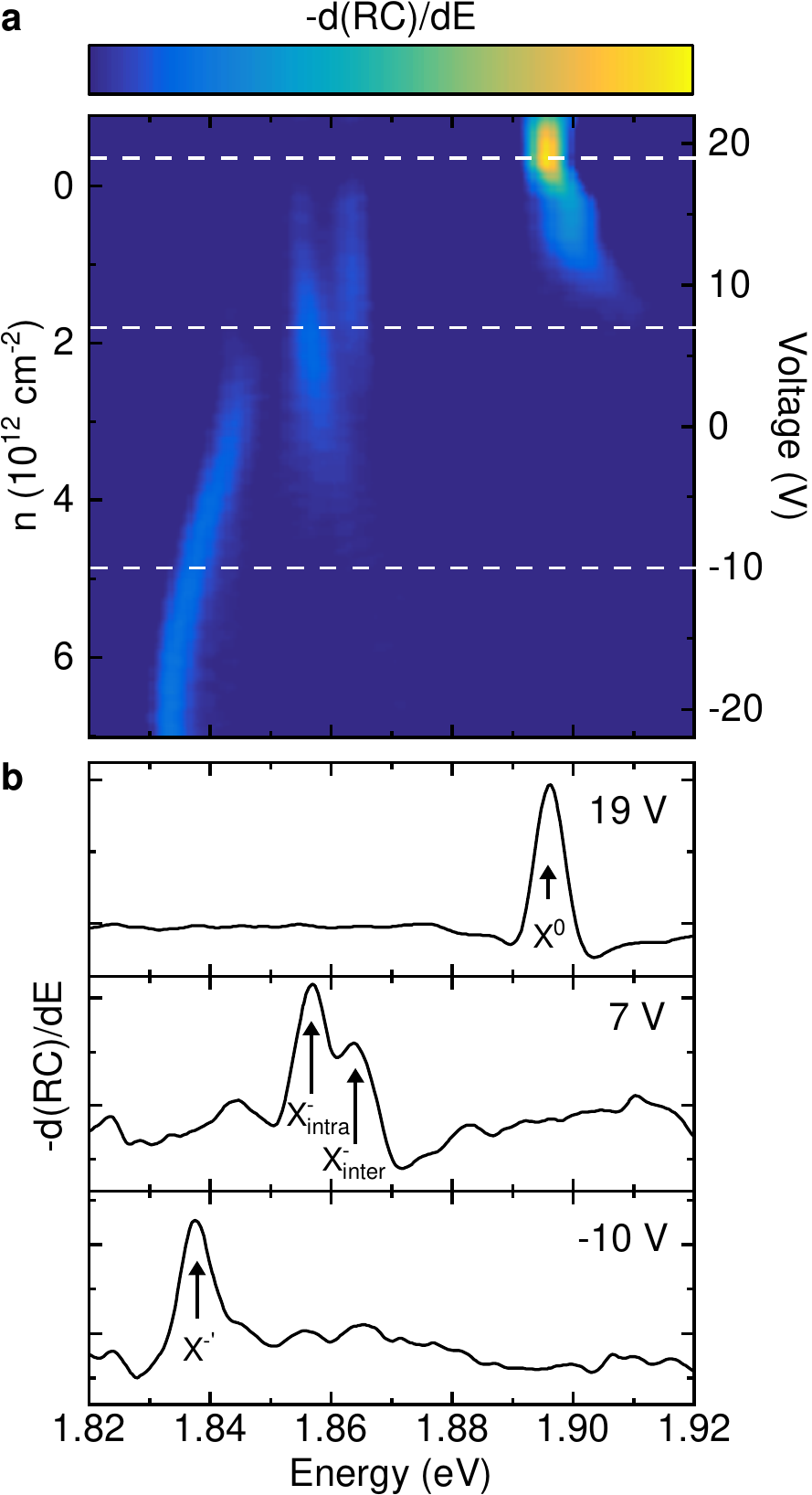}
\caption{Charge dependence of the reflectance contrast spectrum of 1L-W$_\mathrm{Se}^\mathrm{S}$.  \textbf{(a)} RC derivative with varying electron doping density $n$ (left axis) and  applied voltage (right axis) at 4~K. \textbf{(b)} RC derivative spectra at the voltages corresponding to the dashed white linecuts in panel \textbf{a} at 19~V, 7~V and \mbox{-}10~V. The excitonic transitions X$^0$, X$^-_\textrm{inter}$, X$^-_\textrm{intra}$ and X$^{-\prime}$ are labelled.}
\label{fig:3}
\end{figure}

Figure~\ref{fig:2}b shows a RC spectrum from our 1L-W$_\mathrm{Se}^\mathrm{S}$ device (black curve) and the PL spectrum from the same location (red curve).  The RC signal shows a strong feature at 1.893~eV, which confirms our assignment of X$^0$. The lowest observed PL FWHM of the Janus X$^0$ transition is 5.9~meV in our device, the lowest reported to date. The X$^0$ transition is present in both PL and RC across the fully converted Janus region (see S3), with an average PL transition energy of 1.890(1)~eV and an average FWHM of 8.4(4)~meV.

Power-dependent PL measurements (see S4) provide further evidence that X$^0$ is the neutral exciton transition as its intensity scales linearly with power over the measured range 15~nW to 50~$\mu$W (corresponding to 3 Wcm$^{-2}$ to 10$^4$ Wcm$^{-2}$). We note that in the spectral range 1.750 to 1.825 eV we also observe PL peaks with linear power dependences at low power and that saturate in the range 50 to 500 nW (10 to 100 Wcm$^{-2}$). This power saturation suggests the presence of localised defects displaying 
quantum light emission~\cite{kurtsiefer2000stable,he2015single, montblanch2021confinement}.

Density functional theory (DFT) calculations of the 1L-W$_\mathrm{Se}^\mathrm{S}$ band structure (see S5) show that, similar to conventional tungsten-based TMDs (1L-WSe$_2$ and 1L-WS$_2$)~\cite{liu2013three, kosmider2013large,kormanyos2015k}, 1L-W$_\mathrm{Se}^\mathrm{S}$ is direct-bandgap at the K points, with a spin ordering such that the upper valence band is opposite in spin to the lower spin-split conduction band. The spin ordering in the conduction band allows for both a negatively charged intervalley trion (X$^-_\textrm{inter}$), with the two electrons in different valleys, and an intravalley trion (X$^-_\textrm{intra}$), with the two electrons in the same valley. Our DFT calculations predict the transition energies of the Coulomb-exchange split X$^-_\textrm{inter}$ and  X$^-_\textrm{intra}$ to be 26~meV and 32~meV, respectively, below the transition energy of the neutral exciton in free-standing 1L-W$_\mathrm{Se}^\mathrm{S}$.

\section*{Voltage-controlled generation of charged excitons}

To measure the charged excitonic transitions of 1L-W$_\mathrm{Se}^\mathrm{S}$, we tune its doping by applying a voltage $V$ between the 1L-W$_\mathrm{Se}^\mathrm{S}$ and the Si substrate. Figure~\ref{fig:3}a shows the RC derivative signal as we vary the doping density, $n$ (Methods). Similar doping dependence is observed on a second device (see S2). In the operational range of voltages, only the $n$-doped regime is accessible, due to an intrinsic $n$-doping of $\sim 3 \times 10^{12}$~cm$^{-2}$. The previously identified X$^0$ transition, here at 1.896~eV, dominates the RC signal between +21 to +17 V, corresponding to charge neutrality. As we decrease the voltage, and $n$-dope the 1L-W$_\mathrm{Se}^\mathrm{S}$, lower energy transitions appear, which are analogous to the transitions observed in the $n$-doped regime for WSe$_2$~\cite{courtade2017charged,wang2017probing,wang2017valley}.

Figure~\ref{fig:3}b presents the RC derivative at 19, 7 and $-10$~V. The neutral exciton, X$^0$, is shown in the line cut at 19~V. Between +17 to +5~V, we see a doublet, which we identify as X$^-_\textrm{inter}$ and X$^-_\textrm{intra}$ in the line cut at 7~V with peaks at 1.864~eV and 1.857~eV (32~meV and 39~meV below X$^0$ respectively). We attribute the difference in energies of these trions compared to the DFT calculation to a difference in dielectric environment caused by ML-hBN encapsulation. The exchange splitting between the negative trion transitions of 7~meV is in good agreement with our calculations. 

\begin{figure*}
\centering
\includegraphics[width=17.2cm]{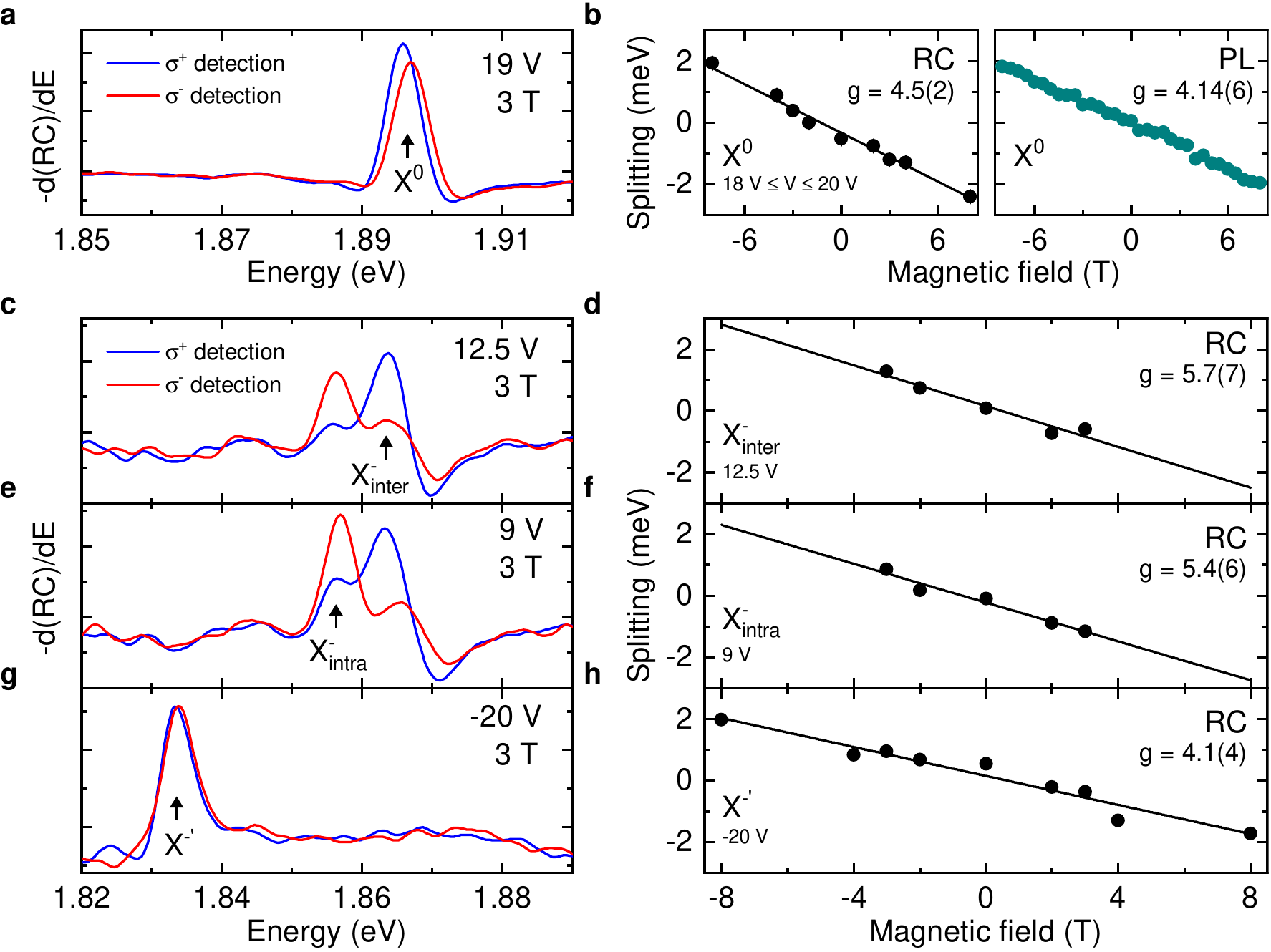}
\caption{Magnetic field dependence of the excitonic complexes in 1L-W$_\mathrm{Se}^\mathrm{S}$. \textbf{(a)}  RC derivative spectrum with $\sigma^+$ (blue) and $\sigma^-$ (red) polarised collection at  $B = 3$~T for X$^0$ (at 19 V). \textbf{(b)} The energy splitting $\Delta E$ between X$^0$ peaks in $\sigma^+$ and $\sigma^-$ detected light as a function of magnetic field. The left panel shows the average splitting measured with RC in the neutral regime (splitting averaged between 18 to 20 V at each magnetic field). The right panel shows the splitting of X$^0$ measured with PL. The solid curve is a linear fit to $\Delta E = - g \mu_\textrm{B} B$, and the g factors are displayed for both RC and PL. \textbf{(c)} Same measurement as in \textbf{a} but for $X^-_\textrm{inter}$ at 12.5 V. \textbf{(d)} $\Delta E$  as a function of magnetic field  for $X^-_\textrm{inter}$ at 12.5 V.  \textbf{(e), (f)} Same as in \textbf{c} and \textbf{d} but for $X^-_\textrm{intra}$ at 9 V. \textbf{(g), (h)} Same as in \textbf{c} and \textbf{d} but for the  X$^{-\prime}$ peak at -20 V. All measurements were carried out at 4~K.}
\label{fig:4}
\end{figure*}

At increased $n$-doping, below 5~V,  the X$^-_\textrm{inter}$ and X$^-_\textrm{intra}$ peaks vanish and a single peak, labelled X$^{-\prime}$ in the linecut at -10~V in Fig.~\ref{fig:3}b, dominates the derivative of the RC spectrum. The X$^{-\prime}$ peak initially appears at 1.845~eV and redshifts by 10~meV between +5 and -17~V. An excitonic transition with a similar doping dependence has previously been observed in 1L-WSe$_2$~\cite{wang2017probing,wang2017valley, Tuan2019} and  attributed to neutral excitons bound to intervalley plasmons \cite{van2017,Tuan2019}. We expect this peak in  1L-W$_\mathrm{Se}^\mathrm{S}$ to be similar in origin.

\section*{Magnetic-field dependence of Janus excitons}

We next probe the exciton g factors by applying an out-of-plane magnetic field, $B$, and measuring the Zeeman energy splitting of the exciton transitions. We send unpolarised light to the device and detect the RC spectra with both $\sigma^+$ and $\sigma^-$ circular polarisations. The left-aligned panels (a, c, e and g) in Fig.~\ref{fig:4} display the RC derivative spectra for each excitonic transition measured at $B = $~3~T magnetic field, with the right-circular  ($\sigma^+$)  and left-circular  ($\sigma^-$) polarisations shown by the blue and red curves, respectively. The splitting $\Delta E$ as a function of $B$ is shown in the right-hand panels (b, d, f and h) of Fig.~\ref{fig:4}. Linear fits give the magnitude of the exciton transition g factors, where $\Delta E = - g\mu_\textrm{B} B$ ($\mu_\textrm{B}$~=~58~$\mu$eV~T$^{-1}$ is the Bohr magneton).

Figure~\ref{fig:4}a presents the RC derivative spectra for X$^0$ at 3~T, showing a well-resolved splitting. Figure~\ref{fig:4}b shows $\Delta E$ for X$^0$ as a function of magnetic field, for both RC and PL. From the linear fit we extract similar g factors of 4.5(2) and  4.14(6) for RC and PL, respectively. For conventional TMDs, g factors with values of $\sim$4 have typically been assigned to bright excitons in the K and K' valleys, with valley, orbital and spin contributing to the magnetic moment~\cite{aivazian2015magnetic,Koperski2018}. The measured g factors are consistent with 1L-W$_\mathrm{Se}^\mathrm{S}$ having a direct-bandgap at the K points.

The g factors of the negatively charged trions depend strongly on doping, ranging from 3-13 for voltages from 8 to 14 V (see S6). For conventional TMDs, this dependence has been attributed to many-body interactions with the Fermi sea of electrons~\cite{Klein2021, lyons2019valley}. Figure~\ref{fig:4}c-f shows RC derivative spectra and splittings as a function of magnetic field for the negative trions at example voltages. We find a g factor of 5.7(7) for X$^-_\textrm{inter}$ and 5.4(6) for X$^-_\textrm{intra}$ at the voltages presented. The X$^-_\textrm{inter}$ and X$^-_\textrm{intra}$ transitions additionally show evidence of the thermalisation of the excess charge, as observed in conventional tungsten-based TMDs~ \cite{Koperski2018, barbone2018charge, Kapuscinski2020}. Beyond $\sim$~3~T, this  leads to only a single polarisation being observable for each negative trion.

Figure~\ref{fig:4}g shows the polarisation-resolved RC derivative spectrum for the X$^{-\prime}$ transition at 3~T. Fig.~\ref{fig:4}h displays the RC splitting of X$^{-\prime}$ as a function of magnetic field, which gives a g factor of 4.1(4), consistent with the interpretation of  X$^{-\prime}$ as the neutral exciton dressed by many-body interactions~\cite{wang2017probing,wang2017valley, Tuan2019}.

\section*{Conclusions}

We have identified  several excitonic complexes in Janus 1L-W$_\mathrm{Se}^\mathrm{S}$: X$^0$, X$^-_\textrm{inter}$, X$^-_\textrm{intra}$ and X$^{-\prime}$ and measured their g factors by integrating a hBN encapsulated 1L-W$_\mathrm{Se}^\mathrm{S}$ into a charge-control device. Integrating J-TMDs into vertical heterostructures is key to designing photovoltaics~\cite{liu2020photogenerated, tang20222d}, while resolving few-meV exciton linewidths and identifying the exciton spectrum determines the suitability of J-TMDs for sensing~\cite{yin2021recent, zhang2022janus}. Future work includes identifying the transitions that give rise to the as-yet unidentified PL peaks as well as measuring the excitonic spectrum in the positively doped regime. An immediate next step is measuring the out-of-plane electric dipole moment of excitons in 1L-W$_\mathrm{Se}^\mathrm{S}$ by applying an out-of-plane electric field in a capacitor-like device structure. The predicted permanent electric dipole moment of 0.24~D~\cite{Li2017} for the Janus X$^0$, means that the resulting Stark shift of 5 meV at 1~V/nm would be resolved with our $\sim$6~meV linewidths.

\section*{Acknowledgements}

We acknowledge funding from the EU Quantum Technology (2D-SIPC) and Graphene Flagships; EU grants CHARM and Graph-X; ERC grants PEGASOS, Hetero2D and GSYNCOR; and EPSRC Grants EP/K01711X/1, EP/K017144/1, EP/N010345/1 and EP/L016087/1. D. M. K. acknowledges support of a Royal Society university research fellowship URF\textbackslash{}R1\textbackslash{}180593. S.T. acknowledges primary support from DOE-SC0020653 (materials synthesis), NSF CMMI 1825594 (NMR and TEM studies), NSF DMR-2206987 (magnetic measurements), NSF CMMI-1933214, NSF 1904716, NSF 1935994, NSF ECCS 2052527, DMR 2111812, and CMMI 2129412 (scalability of Janus layers). 

\appendix

\section{Fabrication}

We build our device by following a multi-step process: first, the bottom ML-hBN is mechanically exfoliated onto a Si/SiO$_2$ (90~nm oxide thickness) substrate. Second, a parent 1L-WSe$_2$ is mechanically exfoliated from a flux-zone grown~\cite{zhang2015flux} bulk WSe$_2$ crystal and deposited on top of the bottom ML-hBN by polydimethylsiloxane (PDMS) transfer. Third, the 1L-WSe$_2$ undergoes AFM flattening~\cite{rosenberger2018nano} and subsequent conversion to a Janus 1L-W$_\mathrm{Se}^\mathrm{S}$ by using the selective epitaxial atomic replacement (SEAR) method~\cite{Tongay2020}, while recording time-resolved Raman spectroscopy measurements in-\textit{situ} to achieve deterministic conversion~\cite{Qin2022}. Fourth, the top ML-hBN and FLG are sequentially deposited on top of the 1L-W$_\mathrm{Se}^\mathrm{S}$ by PDMS transfer, with annealing to 150~$^{\circ}$C and AFM flattening after each layer is deposited. The FLG is mechanically exfoliated from graphite sourced from HQ Graphene. Fifth, gold contacts are deposited using standard electron-beam lithography procedures.

AFM topography (Bruker Icon) is used to confirm the layer thicknesses and Raman spectroscopy (Horiba LabRam Evolution) is used to characterise the various constituents of the heterostructure, along with confirming the conversion from 1L-WSe$_2$ to 1L-W$_\mathrm{Se}^\mathrm{S}$ (see S1).

\section{Optical measurements}

All 4~K measurements were taken in a closed-cycle cryostat (AttoDRY 1000 from Attocube Systems AG), equipped with an 8~T superconducting magnet.

Excitation and collection light pass through a home-built confocal microscope in reflection geometry, with a 0.81 numerical aperture apochromatic objective (LT-APO/NIR/0.81 from Attocube Systems AG). The PL measurements use continuous-wave excitation from a 2.33~eV laser (Ventus 532 from Laser Quantum Ltd.), with the  reported excitation powers measured on the sample and the optical intensity calculated from the optical spot size given by the 0.81 NA. The PL signal is sent to a 150-line grating spectrometer (Princeton Instruments Inc.).

The RC measurements use broadband light (Thorlabs mounted LED M660L4, nominal wavelength 660~nm, FWHM 20~nm).  The reflected light is collected in the confocal microscope discussed above and the spectra are recorded on the same 150-line grating spectrometer as for PL. RC is calculated by comparing the spectrum reflected from the heterostructure in a region with the 1L-W$_\mathrm{Se}^\mathrm{S}$, $R$, and without 1L-W$_\mathrm{Se}^\mathrm{S}$, $R_0$. RC as a function of emission energy $E$ is then given by
\begin{equation*}
    RC(E) = \frac{R(E) - R_0(E)}{R(E)+R_0(E)}.
\end{equation*}

The negative derivative of the RC spectrum, $-\textrm{d}(RC)/\textrm{d}E$, highlights the excitonic transitions and suppresses the RC background~\cite{chernikov2014exciton, glazov2015spin, wang2017probing}. To obtain the derivative RC spectrum, we first smooth the raw RC spectrum using a spline fit and then take the derivative of the resultant spline. 

 \section{Gate-voltage to layer-doping conversion}

The doping density $n$ (charge per unit area) is calculated from the applied voltage $V$ (Keithley 2400 SMU), by using the gate capacitance, $C$ 
\begin{equation*}
    n(V) = n_i - (C V/q_e). 
\end{equation*}
The intrinsic doping, $n_i$, is the doping density at zero applied voltage and the magnitude of the  electron charge is $q_e = 1.6 \times 10^{-19}$~C.

The voltage is applied across both the ML-hBN and  SiO$_2$ and the gate capacitance can be derived by combining the dielectric layers of ML-hBN and SiO$_2$ in series
\begin{equation*}
    C = \epsilon_0 \frac{\epsilon_\textrm{SiO2} \epsilon_\textrm{hBN}}{\epsilon_\textrm{hBN} d_\textrm{SiO2} + \epsilon_\textrm{SiO2} d_\textrm{hBN}}.
\end{equation*}
The  relative dielectric constant of SiO$_2$ and hBN is $\epsilon_\textrm{SiO2}~=~3.9$~\cite{kingon2000alternative} and  $\epsilon_\textrm{hBN}~=~3.8$~\cite{laturia2018dielectric}, respectively. $\epsilon_0 = 8.85 \times 10^{12}$~Fm$^{-1}$ is the vacuum permittivity.  

The  thickness of SiO$_2$ is $d_\textrm{SiO2}~=~90$~nm and the thickness of hBN is $d_\textrm{hBN} = 27$~nm (see S1).  The intrinsic doping density is 
$n_i = 3 \times 10^{12}$~cm$^{-2}$, determined by setting the doping density to $n = 0$ when the reflectance contrast signal from the neutral exciton vanishes (17~V)~\cite{wang2017probing}, where positive $n$ indicates electron doping.

\bibliography{main.bbl}

\end{document}


\title{Supplementary for: \\``Identification of  exciton complexes in a  charge-tuneable  Janus W$_\mathrm{Se}^\mathrm{S}$ monolayer''
}

\author{Matthew S. G. Feuer}
\thanks{These authors contributed equally to this work.}
\affiliation{Cavendish Laboratory, University of Cambridge, 19 J. J. Thomson Ave., Cambridge, CB3 0HE, UK}

\author{Alejandro R.-P. Montblanch}
\thanks{These authors contributed equally to this work.}
\affiliation{Cavendish Laboratory, University of Cambridge, 19 J. J. Thomson Ave., Cambridge, CB3 0HE, UK}

\author{Mohammed Sayyad}
\thanks{These authors contributed equally to this work.}
\affiliation{Materials Science and Engineering, School for Engineering of Matter, Transport and Energy, Arizona State University, Tempe, Arizona, 85287, USA}

\author{Carola M. Purser}
\affiliation{Cavendish Laboratory, University of Cambridge, 19 J. J. Thomson Ave., Cambridge, CB3 0HE, UK}
\affiliation{Cambridge Graphene Centre, University of Cambridge, 9 J. J. Thomson Ave., Cambridge, CB3 0FA, UK}

\author{Ying Qin}
\affiliation{Materials Science and Engineering, School for Engineering of Matter, Transport and Energy, Arizona State University, Tempe, Arizona, 85287, USA}

\author{Evgeny M. Alexeev}
\affiliation{Cambridge Graphene Centre, University of Cambridge, 9 J. J. Thomson Ave., Cambridge, CB3 0FA, UK}
\affiliation{Cavendish Laboratory, University of Cambridge, 19 J. J. Thomson Ave., Cambridge, CB3 0HE, UK}

\author{Alisson R. Cadore}
\affiliation{Cambridge Graphene Centre, University of Cambridge, 9 J. J. Thomson Ave., Cambridge, CB3 0FA, UK}

\author{Barbara L. T. Rosa}
\affiliation{Cambridge Graphene Centre, University of Cambridge, 9 J. J. Thomson Ave., Cambridge, CB3 0FA, UK}

\author{James Kerfoot}
\affiliation{Cambridge Graphene Centre, University of Cambridge, 9 J. J. Thomson Ave., Cambridge, CB3 0FA, UK}

\author{Elaheh Mostaani}
\affiliation{Cambridge Graphene Centre, University of Cambridge, 9 J. J. Thomson Ave., Cambridge, CB3 0FA, UK}

\author{Rados\l{}aw Kal\k{e}ba}
\affiliation{Cavendish Laboratory, University of Cambridge, 19 J. J. Thomson Ave., Cambridge, CB3 0HE, UK}

\author{Pranvera Kolari}
\affiliation{Materials Science and Engineering, School for Engineering of Matter, Transport and Energy, Arizona State University, Tempe, Arizona, 85287, USA}

\author{Jan Kopaczek}
\affiliation{Materials Science and Engineering, School for Engineering of Matter, Transport and Energy, Arizona State University, Tempe, Arizona, 85287, USA}

\author{Kenji Watanabe}
\affiliation{National Institute for Materials Science, Tsukuba, Ibaraki 305-0034, Japan}

\author{Takashi Taniguchi}
\affiliation{National Institute for Materials Science, Tsukuba, Ibaraki 305-0034, Japan}

\author{Andrea C. Ferrari}
\affiliation{Cambridge Graphene Centre, University of Cambridge, 9 J. J. Thomson Ave., Cambridge, CB3 0FA, UK}

\author{Dhiren M. Kara}
\affiliation{Cavendish Laboratory, University of Cambridge, 19 J. J. Thomson Ave., Cambridge, CB3 0HE, UK}

\author{Sefaattin Tongay}

\email{sefaattin.tongay@asu.edu}
\affiliation{Materials Science and Engineering, School for Engineering of Matter, Transport and Energy, Arizona State University, Tempe, Arizona, 85287, USA}

\author{Mete Atat\"{u}re}

\email{ma424@cam.ac.uk}
\affiliation{Cavendish Laboratory, University of Cambridge, 19 J. J. Thomson Ave., Cambridge, CB3 0HE, UK}

\maketitle

\tableofcontents

\section{Device characterisation}
\label{sec:characterisation}

\begin{figure*}
\centering
\includegraphics[width=17.2cm]{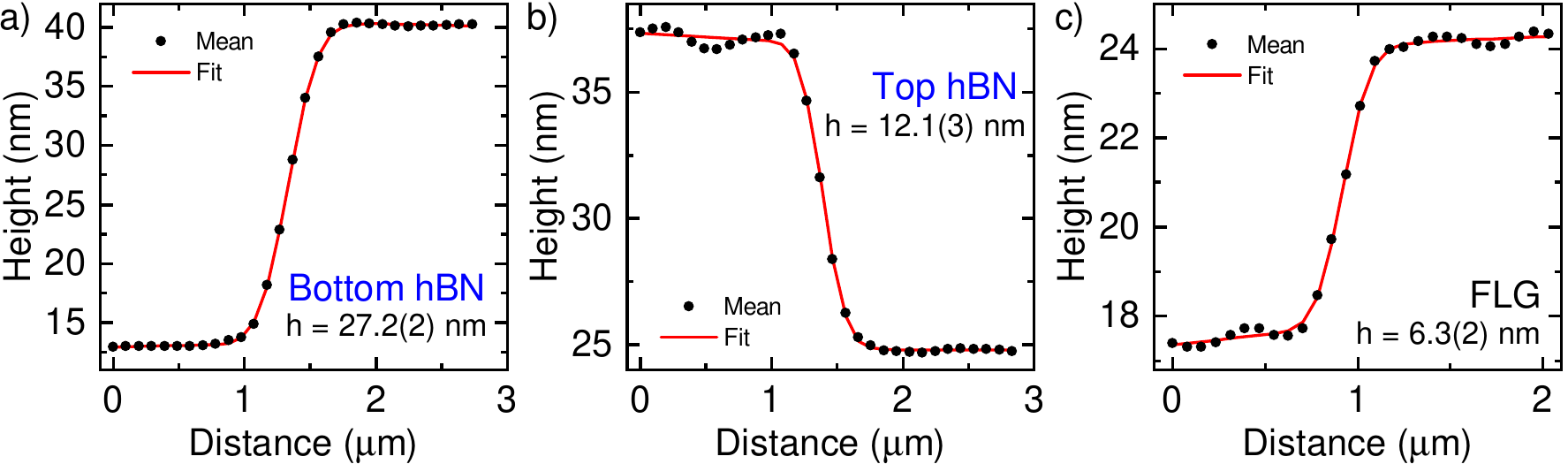}
\caption{{AFM height profiles of the device.}  \textbf{(a)}  Height profile for the bottom ML-hBN, obtained as the average height (black circles) across a line cut. The red line is a fit to a `smooth bent step', which gives the step height $h$. \textbf{(b)} Height profile for the top ML-hBN, obtained as the average height (black circles) across a line cut. The red line is a fit to a `smooth bent step', which gives the step height $h$. \textbf{(c)} Height profile for the FLG, obtained as the average height (black circles) across a line cut. The red line is a fit to a `smooth bent step', which gives the step height $h$. }
\label{fig:AFM}
\end{figure*}

Supplementary Figure~\ref{fig:AFM} shows the characterisation of the layer thicknesses by  atomic force microscope (AFM) topography (Bruker Icon).  Supplementary Fig.~\ref{fig:AFM}a and b show the height profiles for the bottom and top multilayer hexagonal boron nitride (ML-hBN), respectively. Step height analysis was performed in Gwyddion, by obtaining the average height across an appropriate line cut (black circles) and fitting the `smooth bent step' function to extract the layer thickness from the step height~\cite{nevcas2012gwyddion}. This gives the bottom and top ML-hBN thickness as $27.2(2)$~nm and $12.1(3)$~nm respectively, as shown by the red lines in Supplementary Figs.~\ref{fig:AFM}b and \ref{fig:AFM}c. Supplementary Fig.~\ref{fig:AFM}c shows the height profile for the few layer graphene (FLG), which gives the FLG thickness as $6.3(2)$ nm.

\begin{figure*}
\centering
\includegraphics[width=17.2cm]{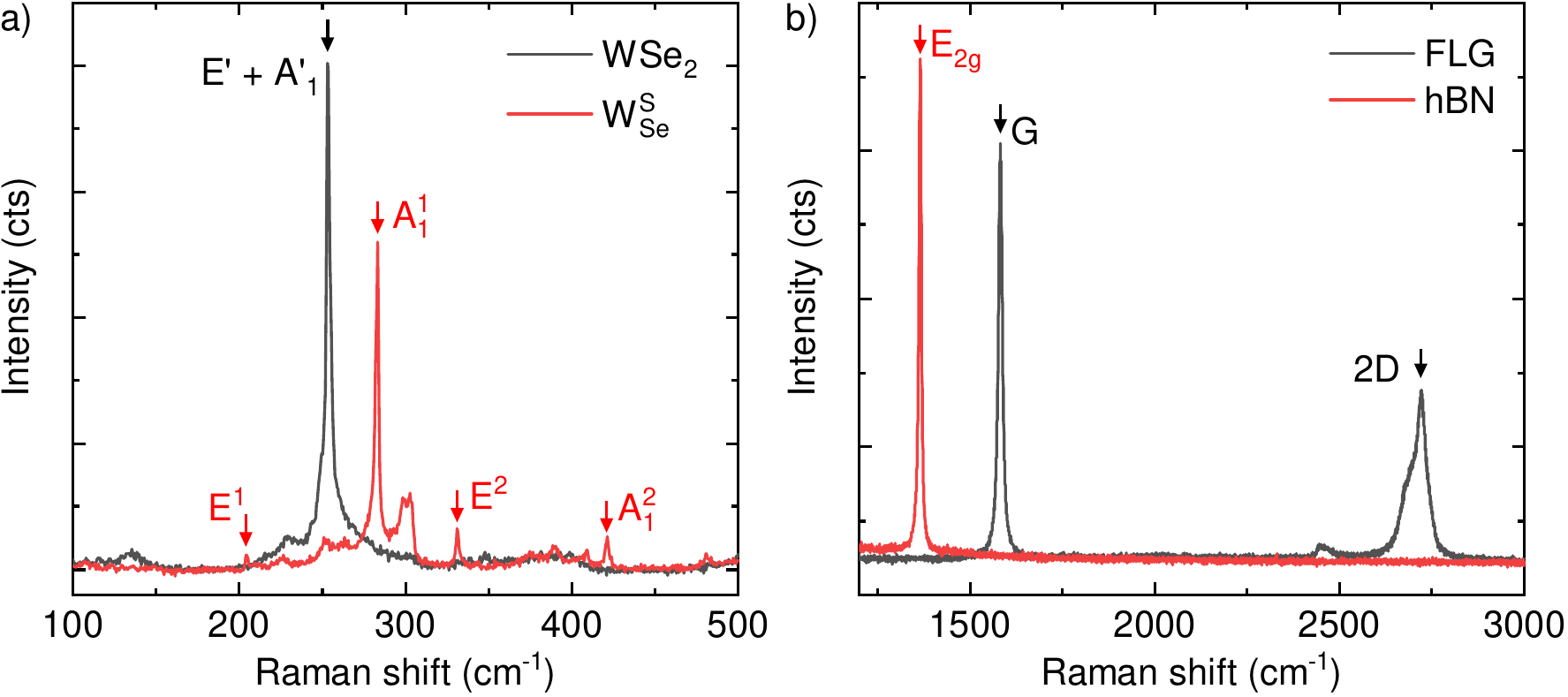}
\caption{{Raman spectra from the device, acquired at room temperature using 2.33~eV optical excitation.} \textbf{(a)} Raman spectra from an un-converted 1L-WSe$_2$ (black curve) and fully converted Janus 1L-W$_\mathrm{Se}^\mathrm{S}$ (red curve) region of the device. The labelled peaks are the 1L-WSe$_2$  $E^\prime$ + $A^\prime_1$ and the 1L-W$_\mathrm{Se}^\mathrm{S}$ first-order Raman modes: $E^1$, $A^1_1$, $E^2$ and $A^2_1$. \textbf{(b)} Raman spectra from the FLG (black curve) and ML-hBN (red curve). The labelled peaks are the  G and 2D Raman modes from the FLG and the $E_{2g}$ Raman mode from the ML-hBN.}
\label{fig:raman}
\end{figure*}

We perform room-temperature Raman spectroscopy in a commercial Horiba LabRam Evolution system, using 2.33~eV optical excitation. Supplementary Figure~\ref{fig:raman}a shows the Raman spectra from an un-converted WSe$_2$ monolayer (1L-WSe$_2$) region of the device as the black curve and fully converted Janus W$_\mathrm{Se}^\mathrm{S}$ monolayer (1L-W$_\mathrm{Se}^\mathrm{S}$) region as the red curve. The peak at $\sim253$ cm$^{-1}$ from 1L-WSe$_2$ is assigned to the convoluted $E^\prime$ + $A^\prime_1$ modes, which are degenerate in  1L-WSe$_2$~\cite{zhao2013lattice, terrones2014new}. The first-order Raman peaks from 1L-W$_\mathrm{Se}^\mathrm{S}$ are assigned as the $E^1$ mode at $\sim205$ cm$^{-1}$, the $A^1_1$ mode at $\sim283$ cm$^{-1}$, the $E^2$ mode at $\sim331$ cm$^{-1}$ and the $A^2_1$ mode at $\sim421$ cm$^{-1}$~\cite{petric2021raman}. The other Raman peaks present arise from higher-order Raman modes, as discussed in ref.~\cite{petric2021raman}. The distinct Raman modes between 1L-WSe$_2$ and Janus 1L-W$_\mathrm{Se}^\mathrm{S}$ distinguishes the un-converted 1L-WSe$_2$ and fully converted 1L-W$_\mathrm{Se}^\mathrm{S}$ regions. Furthermore, the 1L-W$_\mathrm{Se}^\mathrm{S}$ Raman spectra also confirms we have ordered Janus 1L-W$_\mathrm{Se}^\mathrm{S}$ rather than a disordered ternary alloy, which would show the representative Raman peaks of 1L-WSe$_2$ and 1L-WS$_2$~\cite{petric2021raman, Tongay2020, duan2016synthesis}.

Supplementary Figure~\ref{fig:raman}b shows the Raman spectra from the FLG (black curve) and ML-hBN (red curve). The peak at $\sim1366$ cm$^{-1}$ from ML-hBN is assigned to the $E_{2g}$ mode~\cite{reich2005resonant,arenal2006raman}. The G peak from the FLG is seen at  $\sim1582$ cm$^{-1}$ and the 2D band at $\sim2720$ cm$^{-1}$~\cite{ferrari2006raman}.

\begin{figure*}
\centering
\includegraphics[width=17.2cm]{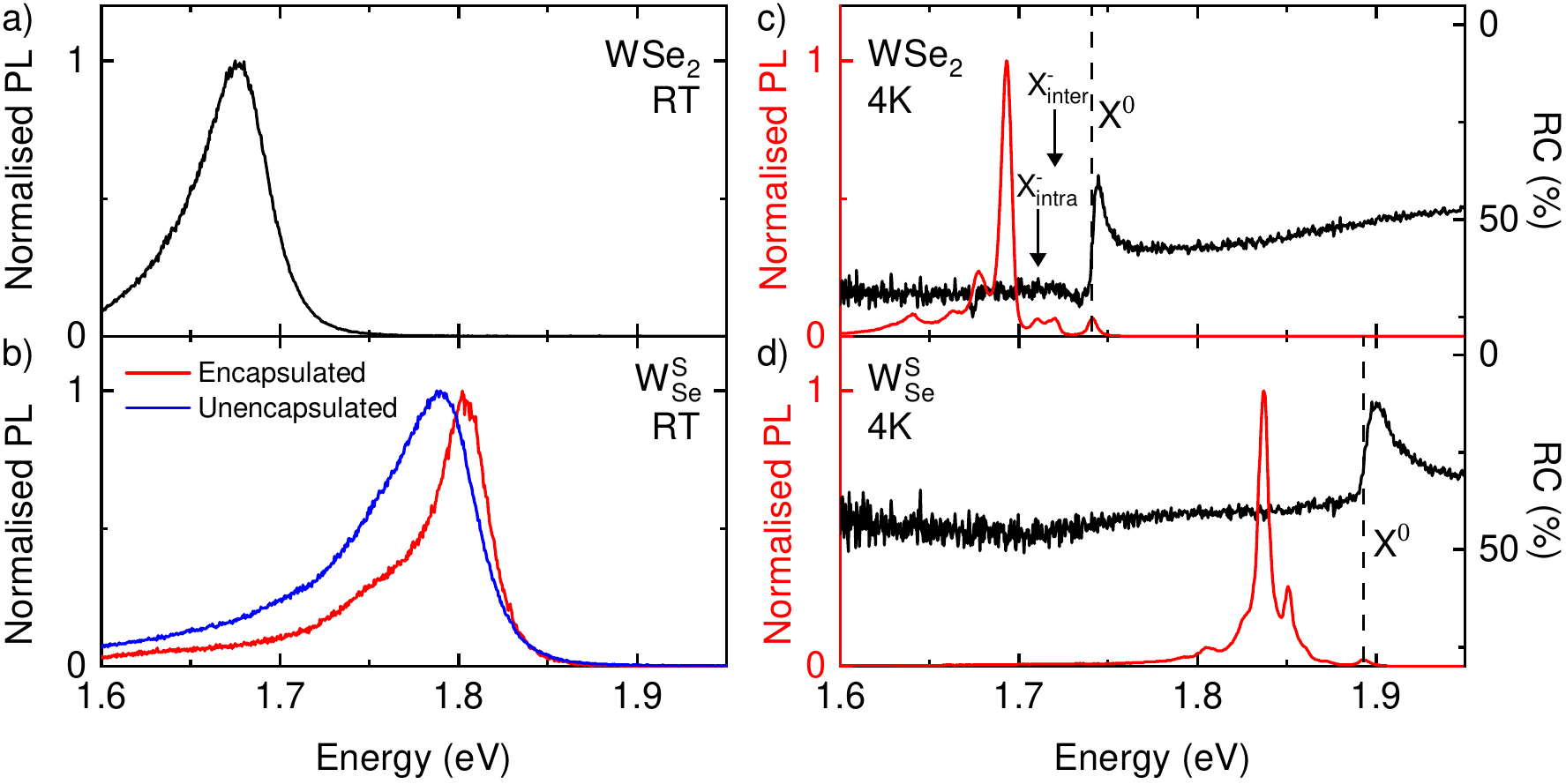}
\caption{{Comparison of spectra from 1L-WSe$_2$ and 1L-W$_\mathrm{Se}^\mathrm{S}$, at room temperature and 4~K.} \textbf{(a)} Room temperature PL spectrum from an un-converted ML-hBN-encapsulated 1L-WSe$_2$ region of the device.  \textbf{(b)}  Room temperature PL spectra from Janus 1L-W$_\mathrm{Se}^\mathrm{S}$, comparing the ML-hBN-encapsulated 1L-W$_\mathrm{Se}^\mathrm{S}$ device (red curve) with unencapsulated 1L-W$_\mathrm{Se}^\mathrm{S}$ (blue curve). Spectra are normalised to the same peak height. \textbf{(c)} 4~K PL (red curve, left axis) and RC (black curve, right axis) spectra from a location in the 1L-WSe$_2$ region of the device. The arrows mark the energy of the intra- and inter-valley trion and the dashed line shows the energy of the neutral exciton, X$^0$.
\textbf{(d)} 4~K PL (red curve, left axis) and RC (black curve, right axis) from a location in the encapsulated1L-W$_\mathrm{Se}^\mathrm{S}$ region of the device. The dashed line marks the energy of the X$^0$. All PL spectra were acquired with 2.33~eV excitation. }
\label{fig:WSe2compare}
\end{figure*}

\begin{figure*}
\centering
\includegraphics[width=17.2cm]{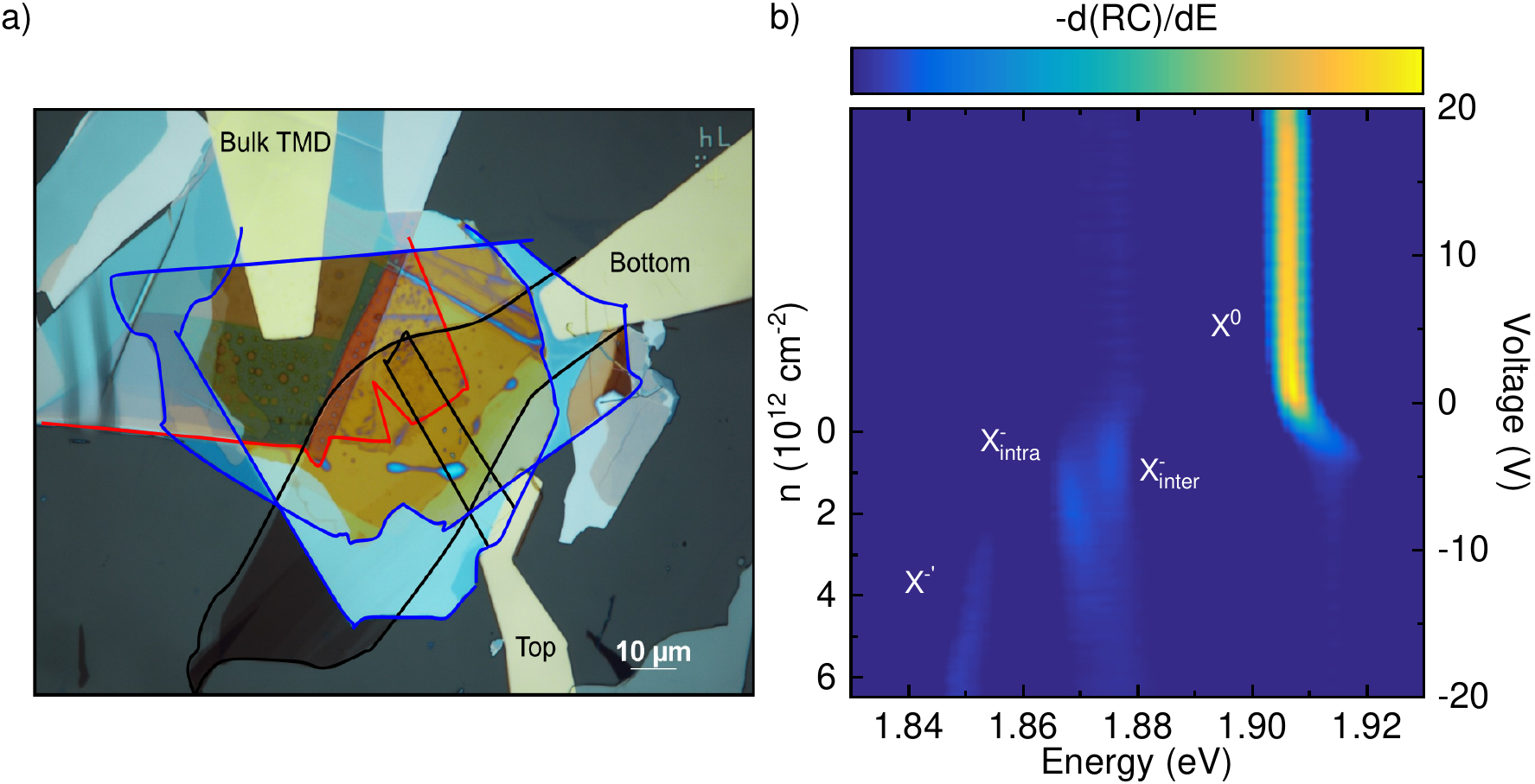}
\caption{{Charge dependence of the reflectance contrast for a second 1L-W$_\mathrm{Se}^\mathrm{S}$ device.} \textbf{(a)} Optical image of the second device. The 1L-W$_\mathrm{Se}^\mathrm{S}$ is outlined in red, the top and bottom ML-hBN in blue and the top and bottom FLG gates in black. The bulk TMD, bottom FLG and top FLG gates are electrically contacted by gold. \textbf{(b)} The derivative of  RC  with varying electron doping density $n$ (left axis) and applied bottom gate voltage (right axis) at 4~K.}
\label{fig:Device2}
\end{figure*}

The photoluminescence (PL) spectra from the 1L-WSe$_2$ and 1L-W$_\mathrm{Se}^\mathrm{S}$ regions of the device are compared in Supplementary Figure~\ref{fig:WSe2compare}. The left-hand panels (a and b) compare the room-temperature PL spectra, with the spectrum from a ML-hBN-encapsulated un-converted 1L-WSe$_2$ region of the device in Supplementary Fig.~\ref{fig:WSe2compare}a. Supplementary Fig.~\ref{fig:WSe2compare}b compares the spectrum from an unencapsulated 1L-W$_\mathrm{Se}^\mathrm{S}$ sample (blue curve), which is directly on a Si/SiO$_2$ substrate and converted from  exfoliated 1L-WSe$_2$, to the ML-hBN-encapsulated 1L-W$_\mathrm{Se}^\mathrm{S}$ (red curve) region of the device. The emission peak intensity is spectrally shifted from  $\sim$1.675~eV in 1L-WSe$_2$ to $\sim$1.8~eV in 1L-W$_\mathrm{Se}^\mathrm{S}$. Peak widths are estimated from a Gaussian fit to the high energy side of the spectra, which gives $\sim$50~meV for the unencapsulated  1L-W$_\mathrm{Se}^\mathrm{S}$, $\sim$42~meV for the encapsulated 1L-WSe$_2$ and $\sim$35~meV for the encapsulated 1L-W$_\mathrm{Se}^\mathrm{S}$. The width of the unencapsulated  1L-W$_\mathrm{Se}^\mathrm{S}$ emission is similar to the best reported for unencapsulated 1L-W$_\mathrm{Se}^\mathrm{S}$ at room temperature~\cite{Qin2022}. Spectral narrowing is seen in the room temperature 1L-W$_\mathrm{Se}^\mathrm{S}$ emission upon encapsulation, to below the width of the emission from the encapsulated 1L-WSe$_2$ region of the device. The highest peak at $\sim 1.8$~eV in the encapsulated 1L-W$_\mathrm{Se}^\mathrm{S}$ spectrum likely arises from the recombination of the neutral exciton, as identified in the temperature dependence of ref.~\cite{Qin2022}.

The PL (red curve) and reflection contrast (RC) spectra (black curve) at 4~K are shown in the right-hand panels (c and d) of Supplementary Figure~\ref{fig:WSe2compare}, for both the 1L-WSe$_2$ region of the device in Supplementary Fig.~\ref{fig:WSe2compare}c and the 1L-W$_\mathrm{Se}^\mathrm{S}$ region in Supplementary Fig.~\ref{fig:WSe2compare}d. The PL spectra narrow considerably from room temperature and multiple resolvable peaks can be seen in both regions. In the 1L-WSe$_2$ region, the PL spectrum is consistent with previous reports~\cite{barbone2018charge} and we can identify the  the neutral exciton, X$^0$, at 1.741~eV (width 5~meV); the inter-valley negatively charged trion, X$^-_\textrm{inter}$, at 1.720~eV (width 6~meV); and the intra-valley negatively charged trion, X$^-_\textrm{intra}$, at 1.710~eV (width 9~meV). The 1L-WSe$_2$ RC spectrum shows the neutral exciton feature at 1.741~eV, coincidental in energy with the X$^0$ PL peak. On moving to the 1L-W$_\mathrm{Se}^\mathrm{S}$ region, the spectral range of the emission shifts to above $\sim$~1.77~eV but the spectral widths remain similar. As discussed in the main text, the highest energy peak is identified as arising from  1L-W$_\mathrm{Se}^\mathrm{S}$ X$^0$ and the RC signal is coincidental with this.

The RC signal from 1L-WSe$_2$ uses a different light source (Thorlabs M730L5) compared to the RC from 1L-W$_\mathrm{Se}^\mathrm{S}$ (Thorlabs M660L4), due to the different spectral range of the transitions.

\section{Gate-dependent measurements in device 2}
\label{sec:Device2}

\begin{figure*}
\centering
\includegraphics[width=17.2cm]{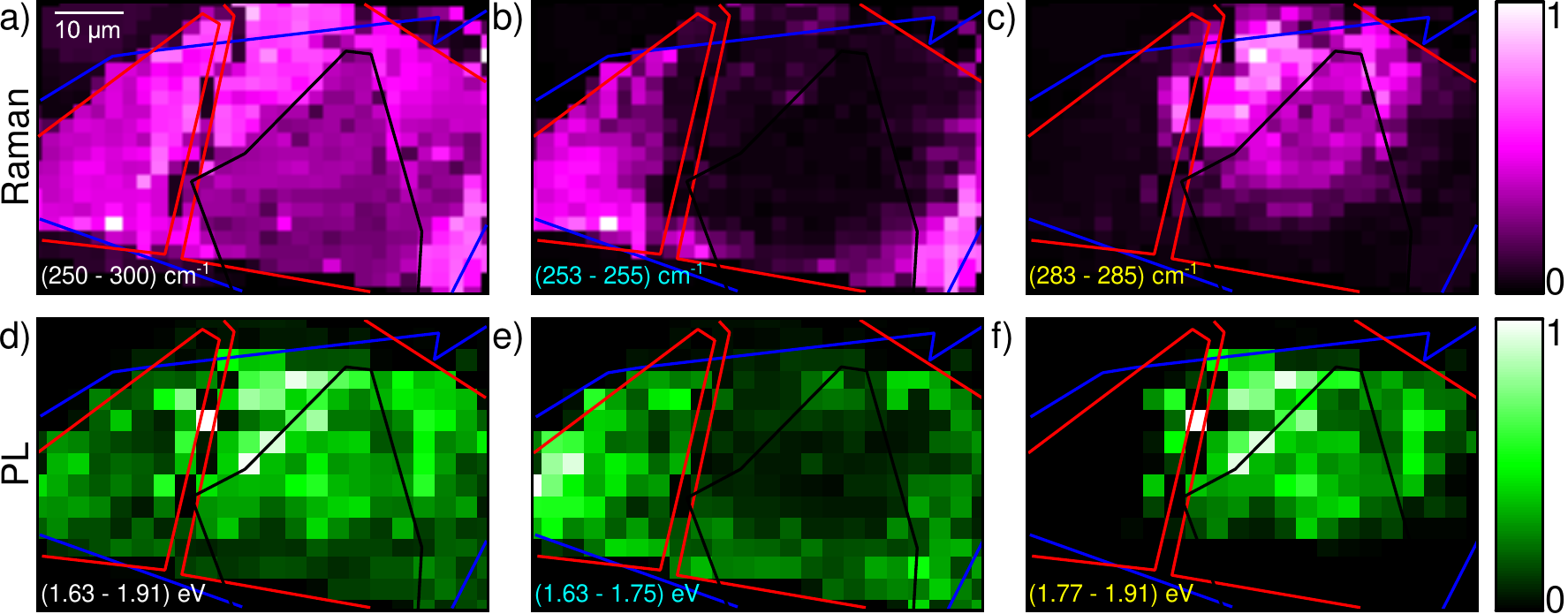}
\caption{{Integrated Raman and PL maps from the device.}  \textbf{(a) - (c)} Integrated Raman maps, acquired at room temperature and using 2.33~eV optical excitation. The integration range is over Raman shifts 250 to 300~cm$^{-1}$ in \textbf{a}, 253 to 255~cm$^{-1}$ (1L-WSe$_2$  $E^\prime$ + $A^\prime_1$ Raman mode) in \textbf{b} and 283 to 285~cm$^{-1}$ (1L-W$_\mathrm{Se}^\mathrm{S}$ $A^1_1$ Raman mode) in \textbf{c}. \textbf{(d) - (f)} Integrated PL maps, acquired at 4~K and using 2.33~eV excitation. The integration range is over the spectral band 1.63 to 1.91~eV in \textbf{d}, 1.63 to 1.75~eV (1L-WSe$_2$ spectral band) in \textbf{e} and 1.77 to 1.91~eV (1L-W$_\mathrm{Se}^\mathrm{S}$ spectral band) in \textbf{f}. All maps are over the same region of the device as highlighted by the white box in Fig.~1b of the main text. Each panel is normalised to the maximum intensity within the panel. The red outline corresponds to the 1L-W$_\mathrm{Se}^\mathrm{S}$, the blue outline to the top ML-hBN and the black outline to the FLG.}
\label{fig:integrated}
\end{figure*}

We have measured similar results to those discussed in the main text on a second sample (device 2). Supplementary Figure~\ref{fig:Device2}a shows the optical image of device 2. This second device consists of a converted region of 1L-W$_\mathrm{Se}^\mathrm{S}$ (outlined in red), attached to bulk unconverted WSe$_2$. The 1L-W$_\mathrm{Se}^\mathrm{S}$ is encapsulated by 61~nm of bottom ML-hBN and 43~nm of top ML-hBN (both outlined in blue). A bottom FLG gate is under the bottom ML-hBN and top FLG gate above the top ML-hBN (both outlined in black). Electrical contacts to the bulk TMD, top and bottom FLG gates are provided by gold leads. The device is fabricated using the same polydimethylsiloxane  (PDMS) transfer technique as for device 1 (see Methods) and the Janus conversion uses the same SEAR in-\textit{situ} process~\cite{Tongay2020, Qin2022}. Layer thicknesses are confirmed by AFM topography.

Supplementary Figure~\ref{fig:Device2}b shows the RC derivative as we tune the doping density, $n$, for device 2.  The doping density is tuned here by applying a voltage between the TMD and bottom FLG gate and $n$ is calculated as described in Methods, using the 61~nm bottom ML-hBN thickness. We set the band edge $n=0$ to be where the neutral exciton RC signal vanishes and find the intrinsic doping to be $n_i = 0.7 \times 10^{12} \textrm{ cm}^{-2}$.  In the \textit{n}-doped regime we see the same transitions and behaviour as for device 1 (see Fig.~3 of the main text), namely the neutral exciton X$^0$, the inter- and intra-valley negative trions (X$^-_\textrm{inter}$ and  X$^-_\textrm{intra}$, respectively)  and the X$^{-\prime}$ transition. Device 2 has a lower level of intrinsic doping than device 1; however, as we change to positive voltage and attempt to enter the \textit{p}-doped regime we observe that the neutral exciton persists and thus we are unable to reach the \textit{p}-doped regime. 

We find that the neutral exciton transition X$^0$ is at 1.907~eV at this location on device 2, which is a 17~meV shift from the average across device 1 and possibly arises from a difference in strain on the device.  We see the X$^-_\textrm{inter}$ transition at 1.875~eV and X$^-_\textrm{intra}$    at 1.869~eV, which gives binding energies relative to X$^0$ of 32~meV and 38~meV, respectively, agreeing with those measured on device 1. This gives an exchange splitting of 6~meV, agreeing with both our calculation (see~\ref{sec:calc}) and measurement on device 1. As in device 1, we see the X$^{-\prime}$ transition, which emerges at 1.854~eV and red shifts with increased doping. The doping densities where the various transitions appear in device 2 are also similar to those in device 1. Specifically, the cross-over between negative trions and X$^{-\prime}$ occurs around $(2-3)\times10^{12}\textrm{cm}^{-2} $ in both devices.

\section{Spectral mapping and homogeneity}
\label{sec:mapping}

\begin{figure*}
\centering
\includegraphics[width=17.2cm]{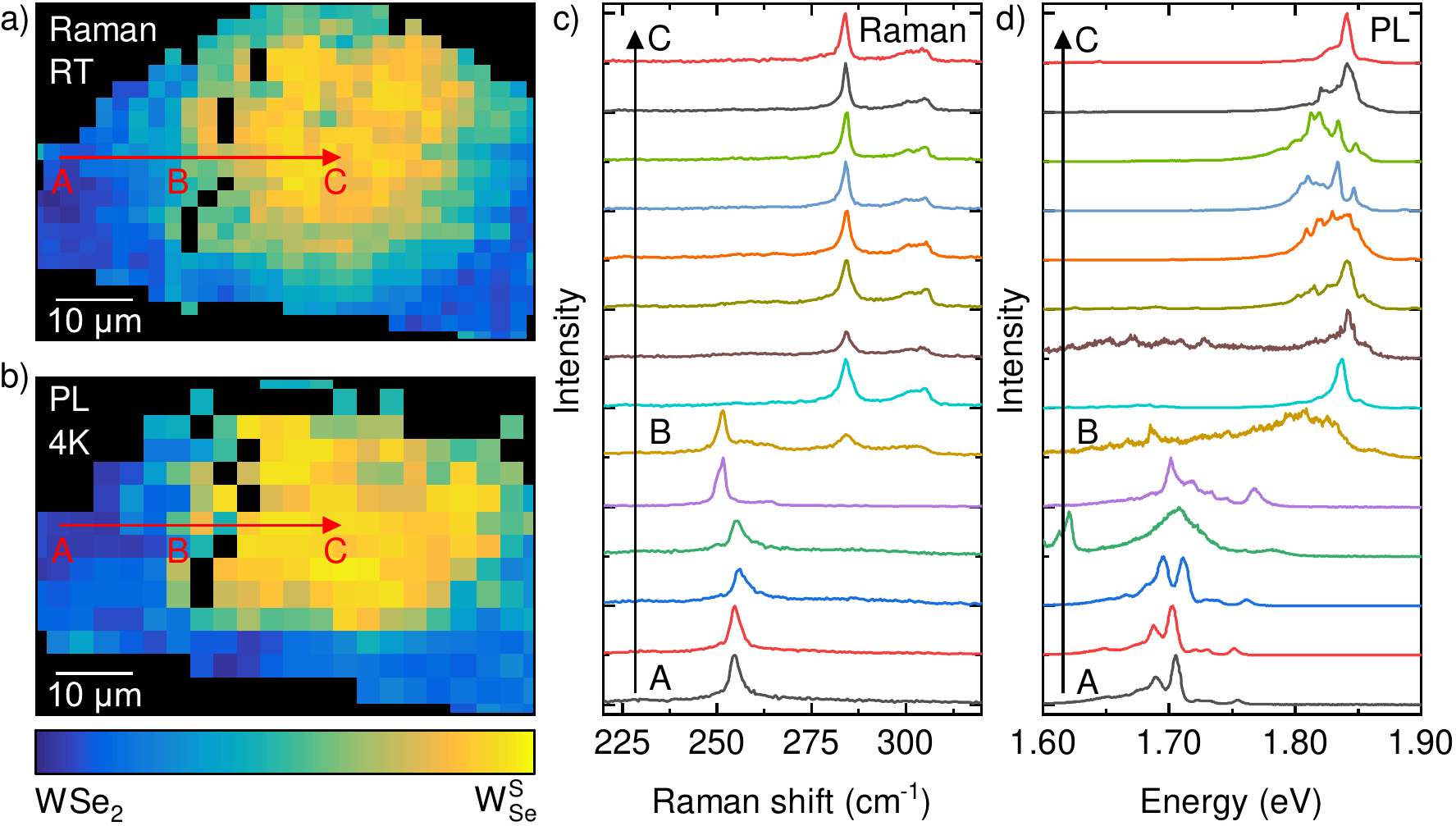}
\caption{{Raman and PL spectra across paths in the Raman and PL maps.} \textbf{(a)} Raman map of the device from Fig.~1c of the main text, acquired at room temperature and using 2.33~eV optical excitation.  \textbf{(b)} PL map of the device from Fig.~1d of the main text, acquired at  4~K and using 2.33~eV excitation. \textbf{(c)} Raman spectra along the red path in \textbf{a}, from the un-converted 1L-WSe$_2$ region (point A), through the partially converted region (point B) to the fully converted 1L-W$_\mathrm{Se}^\mathrm{S}$ region (point C). The spectra are vertically offset for clarity.  \textbf{(d)} PL spectra along the red path in \textbf{b}, from the un-converted 1L-WSe$_2$ region (point A), through the partially converted region (point B) to the fully converted 1L-W$_\mathrm{Se}^\mathrm{S}$ region (point C). The spectra are vertically offset for clarity.}
\label{fig:linecut}
\end{figure*}

Supplementary Figure~\ref{fig:integrated} presents integrated spectral maps over the device, for both Raman and PL spectroscopy, which are used to calculate the relative intensity maps in Figs.~1c and d of the main text. Supplementary Figs.~\ref{fig:integrated}a - \ref{fig:integrated}c show the room-temperature integrated Raman maps of the characteristic Raman modes from 1L-WSe$_2$ and 1L-W$_\mathrm{Se}^\mathrm{S}$. Supplementary Fig.~\ref{fig:integrated}a shows the integrated Raman intensity over the spectral range (250 to 300~cm$^{-1}$) covering both the 1L-WSe$_2$ and 1L-W$_\mathrm{Se}^\mathrm{S}$ characteristic modes. Supplementary Fig.~\ref{fig:integrated}b shows the integrated intensity from the 1L-WSe$_2$  $E^\prime$ + $A^\prime_1$ Raman mode (253 to 255~cm$^{-1}$) and Supplementary Fig.~\ref{fig:integrated}c shows the integrated intensity from the 1L-W$_\mathrm{Se}^\mathrm{S}$ $A^1_1$ Raman mode (283 to 285~cm$^{-1}$). This shows that the flake has a central region of fully converted 1L-W$_\mathrm{Se}^\mathrm{S}$, as shown by the yellow region in the relative intensity map of Fig.~1c from the main text.

Supplementary Figs.~\ref{fig:integrated}d - \ref{fig:integrated}f show the PL maps, acquired at 4 K, of the integrated spectral bands associated with 1L-WSe$_2$ and 1L-W$_\mathrm{Se}^\mathrm{S}$. Supplementary Fig.~\ref{fig:integrated}d shows the integrated PL intensity over the spectral band (1.63 to 1.91~eV) covering both the 1L-WSe$_2$ and 1L-W$_\mathrm{Se}^\mathrm{S}$ emission. Supplementary Figs.~\ref{fig:integrated}e and \ref{fig:integrated}f show the integrated intensity from the 1L-WSe$_2$ (1.63 to 1.75~eV) and 1L-W$_\mathrm{Se}^\mathrm{S}$ (1.77 to 1.91~eV) spectral bands, respectively. The integrated PL maps show a central region of converted 1L-W$_\mathrm{Se}^\mathrm{S}$, consistent with  the integrated Raman maps and shown by the yellow region in the relative intensity map of Fig.~1d from the main text.

Supplementary Figure~\ref{fig:linecut} shows the variation in Raman and PL spectra across paths of varying Janus conversion. Supplementary Figure~\ref{fig:linecut}c and \ref{fig:linecut}d plot the Raman and PL spectra along the paths indicated in Supplementary Figs.~\ref{fig:linecut}a and \ref{fig:linecut}b, respectively. Note that these paths were taken on the same region as the one shown in Figs.~1c and 1d from the main text. The paths are along the red lines from the un-converted  1L-WSe$_2$ region (point A), through the partially converted region (point B), to the fully converted  1L-W$_\mathrm{Se}^\mathrm{S}$ region (point C). The partially converted regions show the distinctive Raman modes from both 1L-WSe$_2$ and 1L-W$_\mathrm{Se}^\mathrm{S}$, along with PL emission in both spectral bands. The un-converted and fully converted regions show the Raman modes corresponding to either   1L-WSe$_2$ or 1L-W$_\mathrm{Se}^\mathrm{S}$, and PL emission from a single spectral band.

\begin{figure*}
\centering
\includegraphics[width=17.2cm]{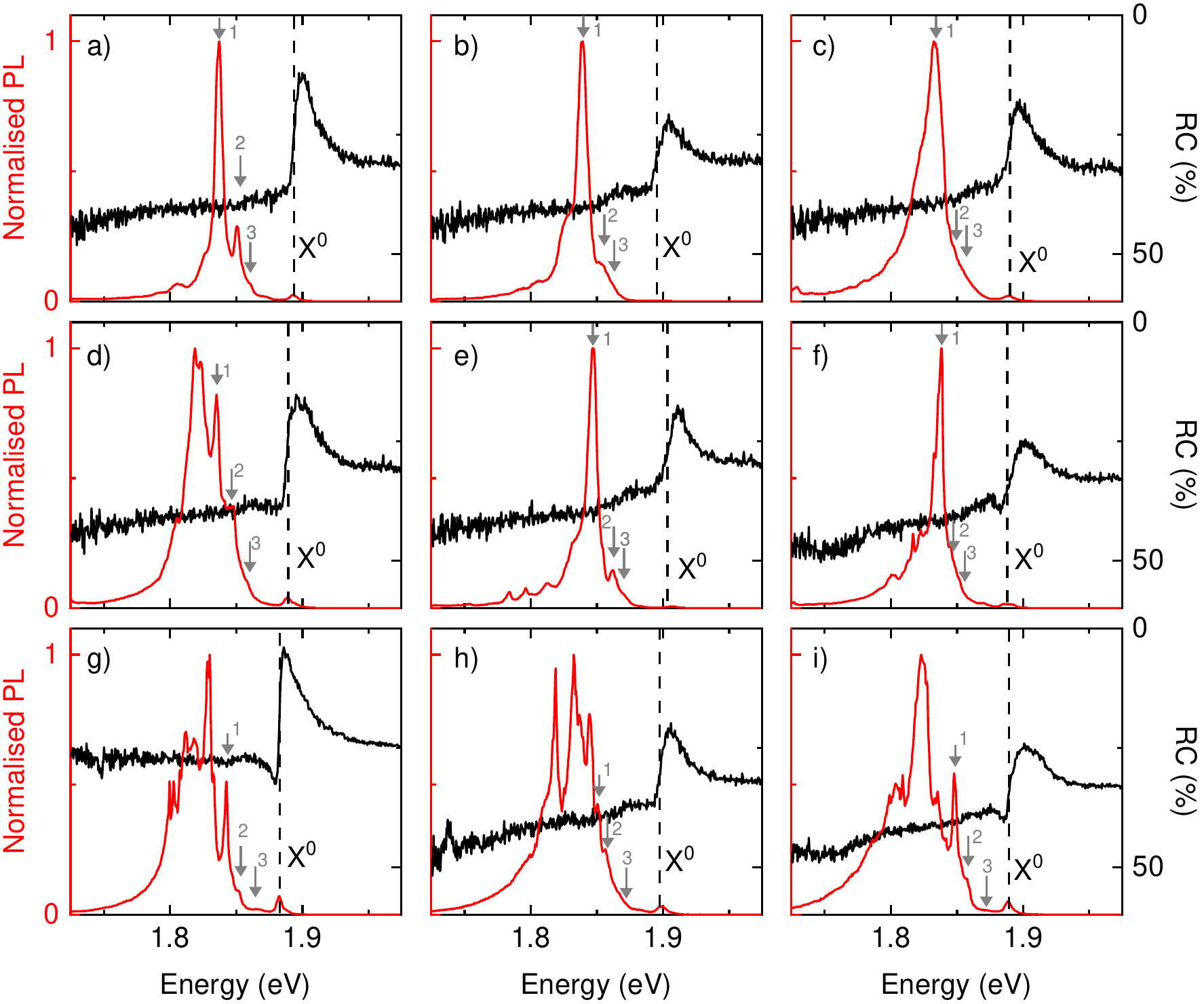}
\caption{{PL and RC spectra from 1L-W$_\mathrm{Se}^\mathrm{S}$ across different locations.}  \textbf{(a) - (i)} PL (red curve, left axis) and RC (black curve, right axis) spectra at different locations in the 1L-W$_\mathrm{Se}^\mathrm{S}$ region of the device. Each panel shows the PL and corresponding RC spectra at a single location. All spectra were acquired at 4~K, and the PL spectra with 2.33~eV excitation. Panel \textbf{a} appears as Fig.~2b of the main text.} 
\label{fig:locations}
\end{figure*}

Supplementary Fig.~\ref{fig:locations} presents the variation in PL and RC spectra at nine locations across the sample in the 1L-W$_\mathrm{Se}^\mathrm{S}$ region of the device. Across the sample, we see changes in the relative intensity of the different PL peaks and width of the redder emission tail. However, the general structure of the higher energy peaks persists across different locations, and specifically the peaks labelled in Fig.~2a of the main text (1, 2, 3 and X$^0$) are identifiable. Across the device, the RC feature is coincidental with the X$^0$ peak in PL but with contrast dependent on location. As discussed in the main text, calculating statistics across the device gives the average X$^0$ PL peak energy as 1.890(1)~eV and FWHM as 8.4(4)~eV. A similar variation in PL spectra is also seen in the best WSe$_2$ samples~\cite{barbone2018charge,rodriguez2021excitons} and likely arises from a variation in strain and fabrication inhomogeneity.

\section{Power-dependent measurements}
\label{sec:power}

\begin{figure*}
\centering
\includegraphics[width=17.2cm]{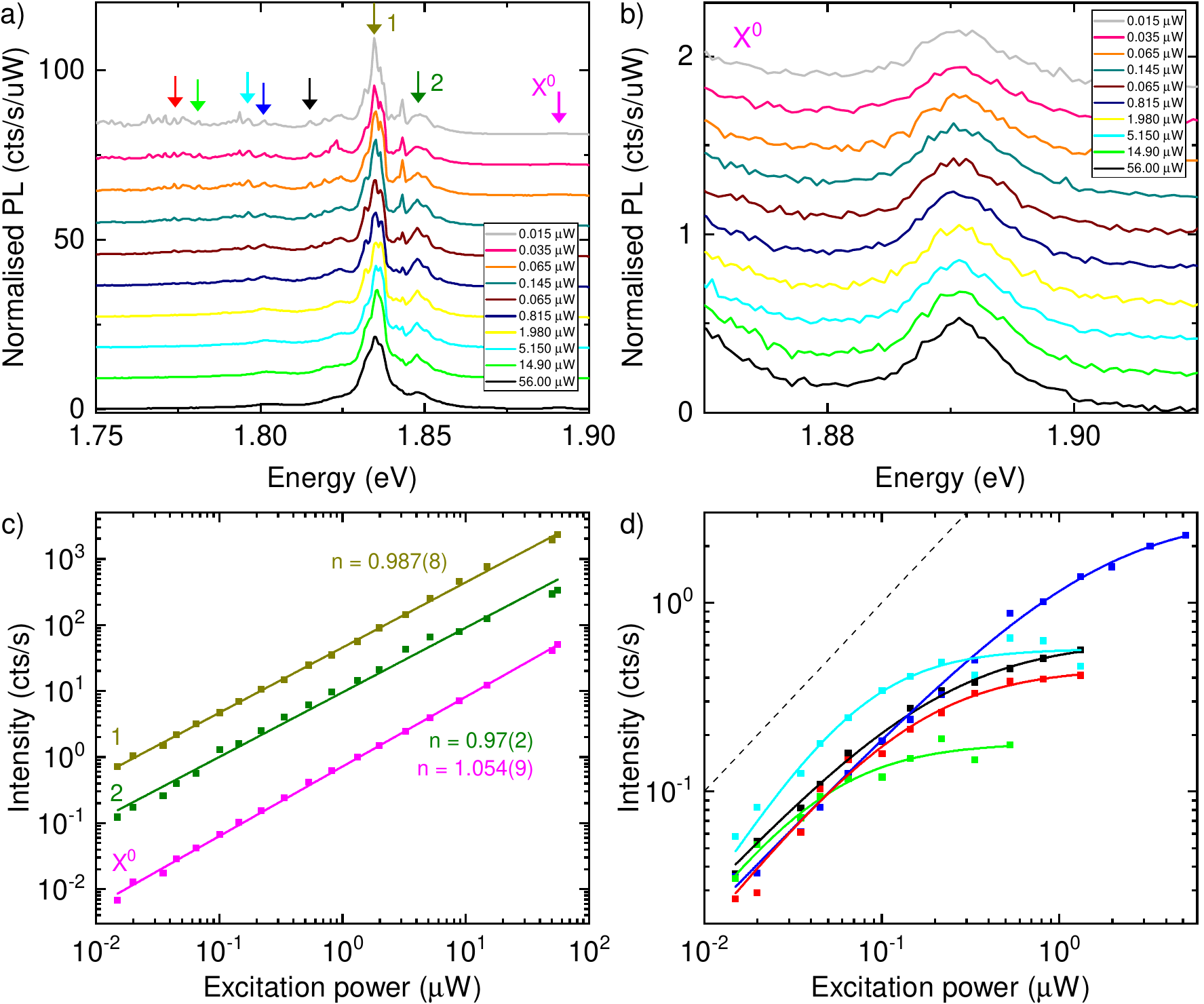}
\caption{{Excitation-power PL dependence of the 1L-W$_\mathrm{Se}^\mathrm{S}$ emission, acquired at 4~K and using 2.33~eV excitation.}  \textbf{(a)} Power dependence of the PL spectra, normalised by integration time and excitation power. The excitation power is varied between 15~nW (grey curve) to 56~$\mu$W (black curve). Spectra are vertically offset for clarity. \textbf{(b)} Same as in \textbf{a}, but showing only the neutral exciton X$^0$. \textbf{(c)} Excitation power dependence of the PL intensity of the labelled peaks 1, 2 and X$^0$ from \textbf{a} and \textbf{b}. These are the resolvable peaks at high power. The data is plotted on a double logarithmic scale and the solid line is a fit to $I~\propto~P^n$, with peak intensity $I$, power $P$ and power law scaling $n$. All three peaks give linear power law scaling. \textbf{(d)} Same as in \textbf{c} but for the unlabelled peaks in \textbf{a} as indicated by coloured arrows, which appear at low excitation power. The solid line is a fit to $I~\propto~P^n/(P^n + P_\textrm{sat}^n)$, with saturation power $P_\textrm{sat}$. The dashed line shows a linear power scaling $n = 1$. All these peaks show an initially linear power dependence, before their intensity saturates.}
\label{fig:power}
\end{figure*}

Supplementary Figure~\ref{fig:power} shows the excitation power dependence of the PL emission from 1L-W$_\mathrm{Se}^\mathrm{S}$, over the full spectral range in Supplementary Fig.~\ref{fig:power}a and over the spectral range of the neutral exciton in Supplementary Fig.~\ref{fig:power}b. The spectra remain of a similar shape as the excitation power is changed from 15 nW to 50 $\mu$W (corresponding to 3 Wcm$^{-2}$ to 10$^4$ Wcm$^{-2}$), except for the appearance of low energy peaks at low powers. The excitation power dependence of the integrated intensity of the resolvable peaks (labelled 1, 2 and X$^0$) in the high power spectrum are shown in  Supplementary Fig.~\ref{fig:power}c. A power law fit shows that these three peaks follow a linear power law scaling, consistent with assignment of these peaks as arising from free excitonic species~\cite{barbone2018charge}. Specifically, the linear power dependence of the X$^0$ peak provides evidence of it arising from the recombination of the neutral exciton, as discussed in the main text. We see no evidence of a blueshift to these spectral peaks with increasing excitation power, up to the maximum accessible power of $\sim 50~\mu$W (10$^4$ Wcm$^{-2}$). The power dependence of the intensity of the marked low energy peaks (labelled by coloured arrows) are shown in Supplementary Fig.~\ref{fig:power}d. In contrast to the higher energy peaks, we see that their intensity follows a linear power scaling at low powers and saturates at powers in the range 50 to 500~nW (10 to 100~Wcm$^{-2}$). As noted in the main text, this points to the presence of localised defects displaying quantum light emission~\cite{kurtsiefer2000stable,he2015single, montblanch2021confinement}.

\section{Band structure calculations}
\label{sec:calc}

\begin{figure*}
\centering
\includegraphics[width=17.2cm]{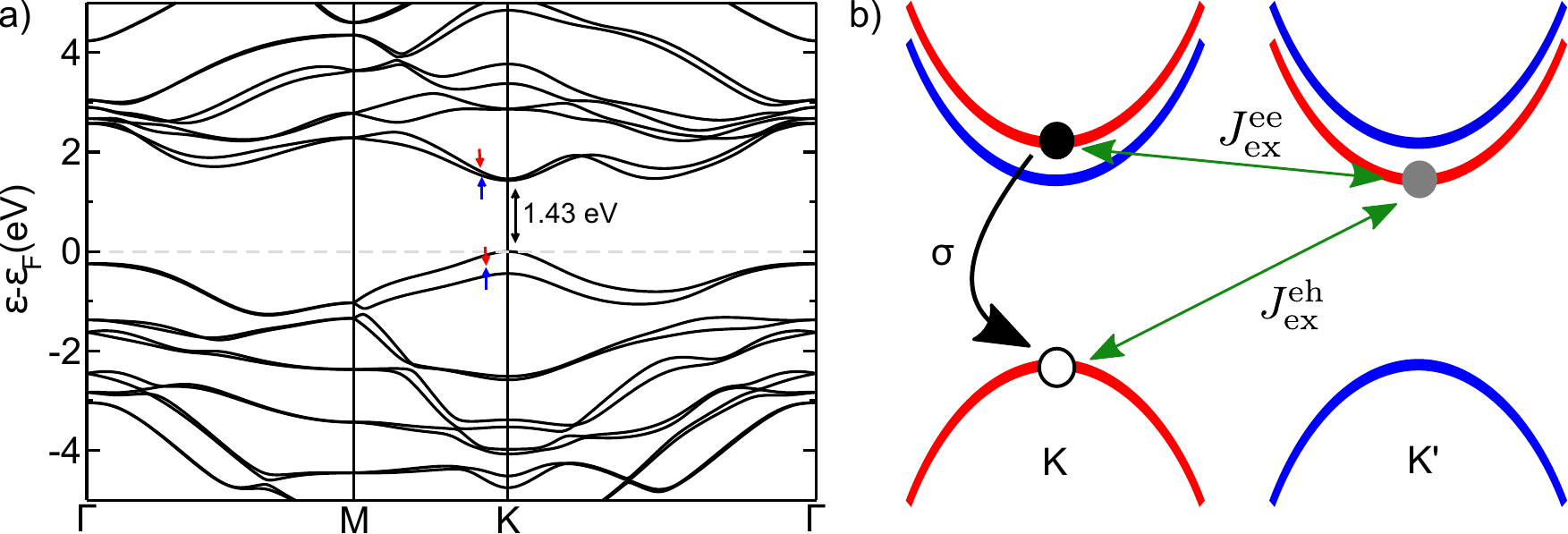}
\caption{{DFT calculation of the band structure for 1L-W$_\mathrm{Se}^\mathrm{S}$.} \textbf{(a)} Energy dispersion plotted along the $\Gamma - M - K - \Gamma$ path. The calculation uses the PBE functional, including spin-orbit coupling. The blue and red arrows indicate the spin of the bands at the K point (spin up and down respectively) and the black arrow shows the energy gap. The Fermi energy is set to zero. \textbf{(b)} The exchange energy for the inter-valley negative trion, between the excess electron (grey circle) and the paired electron-hole pair (black and white circle). $J^\textrm{ee}_\textrm{ex}$ and $J^\textrm{eh}_\textrm{ex}$ are the electron-electron and electron-hole exchange energies, respectively. $\sigma$ indicates the paired electron-hole pair. Spin up bands are shown in blue and spin down in red.}
\label{fig:DFT}
\end{figure*}

\begin{table}
    \centering
    \begin{tabular}{l  c}
        Complex & Binding energy (meV) \\
        \hline
         Negative trion (X$^-$) & 32.2 \\
         Positive trion (X$^+$) & 32.0 \\
         Biexciton (XX$^0$) & 21.0 \\
         Quinton (XX$^-$) & 54.2 \\
         \hline
    \end{tabular}
    \caption{\textbf{Binding energy of the excitonic complexes compared to the neutral exciton, for free-standing 1L-W$_\mathrm{Se}^\mathrm{S}$}.}
    \label{tab:binding}
\end{table}

\begin{figure*}
\centering
\includegraphics[width=17.2cm]{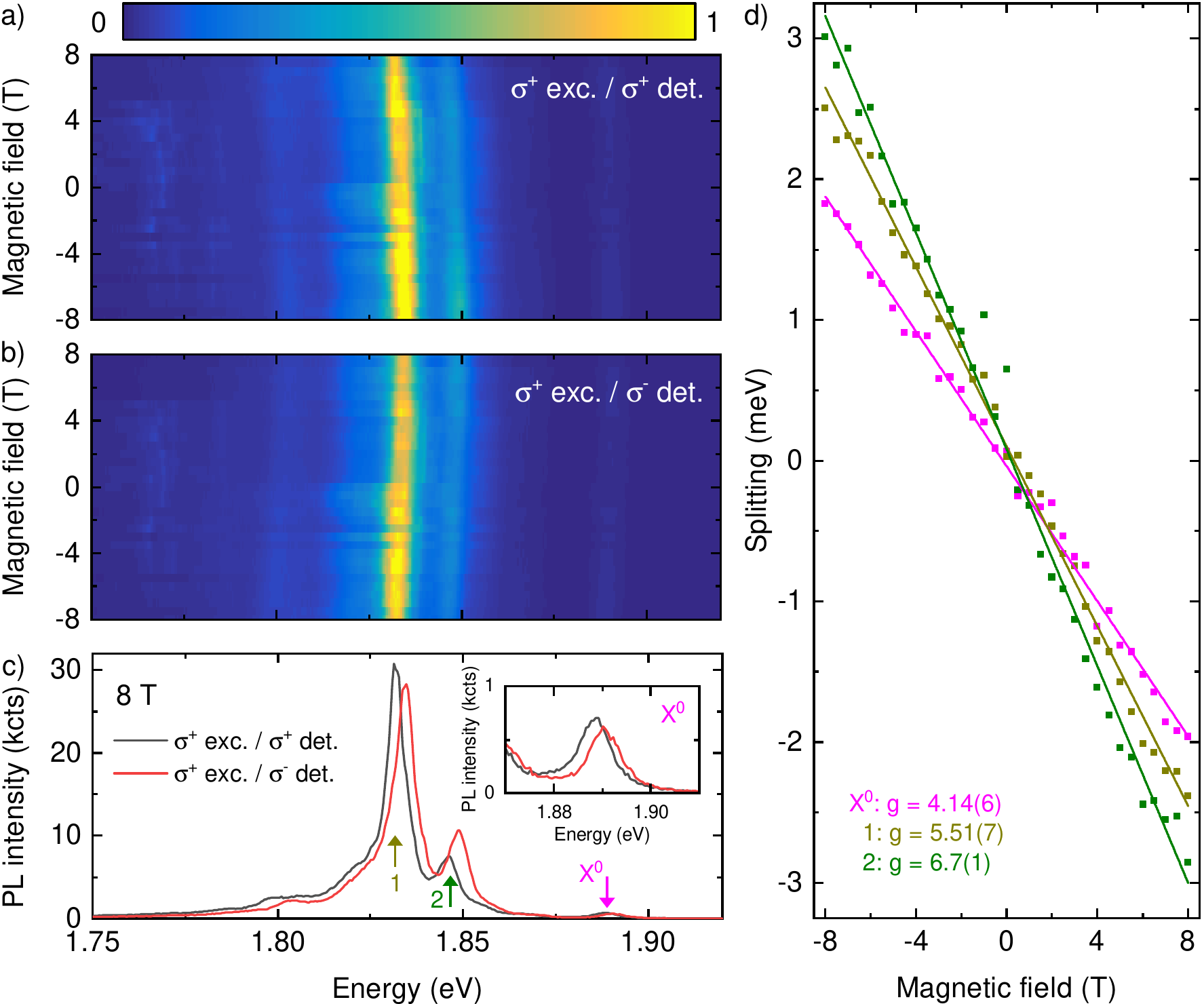}
\caption{{Magnetic field dependence of the PL for 1L-W$_\mathrm{Se}^\mathrm{S}$, acquired at 4~K and using 2.33~eV excitation.}  \textbf{(a)} PL spectra as a function of magnetic field, under right-circularly ($\sigma^+$) polarised excitation and detection (co-polarised). \textbf{(b)} Same as for \textbf{a} but with right-circularly ($\sigma^+$) polarised excitation and left-circularly ($\sigma^-$) polarised  detection (cross-polarised). \textbf{(c)} PL spectra at 8~T, under $\sigma^+$ excitation and with both $\sigma^+$ (black curve) and $\sigma^-$ (red curve) detection. The inset shows only the neutral exciton X$^0$. \textbf{(d)}  Energy splitting $\Delta E$ as a function of  magnetic field $B$ between the  $\sigma^+$ and  $\sigma^-$ detected PL for the labelled peaks in \textbf{c}. The solid lines  are linear fittings, with the g factor found from the gradient as $\Delta E = - g \mu_B  B$. }
\label{fig:magnetic}
\end{figure*}

We perform density functional theory (DFT) to understand the band structure and observed excitonic transitions of 1L-W$_\mathrm{Se}^\mathrm{S}$. Supplementary Figure~\ref{fig:DFT}a shows the band structure along the  $\Gamma - M - K - \Gamma$ path. This is calculated using the DFT-PBE functional, including spin-orbit coupling with a k-grid of $24 \times 24 \times 1$. The DFT calculation shows that, similar to conventional tungsten-based TMDs (1L-WSe$_2$ and 1L-WS$_2$)~\cite{liu2013three, kosmider2013large,kormanyos2015k}, 1L-W$_\mathrm{Se}^\mathrm{S}$ is direct bandgap at the K points with spin ordering such that the upper valence band is opposite in spin to the lower spin-split conduction band, with 37~meV spin splitting.

The effective masses obtained from the DFT calculation are used in the Mott-Wannier model and quantum Monte Carlo (QMC)~\cite{mostaani2017diffusion, barbone2018charge} to calculate the expected binding energies of the excitonic charge complexes in free-standing 1L-W$_\mathrm{Se}^\mathrm{S}$, as given in Supplementary Table~\ref{tab:binding}. Experimental differences arise from the difference in dielectric environment caused by the ML-hBN encapsulation~\cite{hsu2019dielectric}.

The fine structure of the negative trion arises from the exchange interaction between the excess electron and the paired electron-hole, which causes an energy increase for the intervalley relative to intravalley negative trion~\cite{yu2014dirac, plechinger2016trion, courtade2017charged}. Supplementary Figure~\ref{fig:DFT}b shows the electron-electron ($J^\textrm{ee}_\textrm{ex}$) and electron-hole ($J^\textrm{eh}_\textrm{ex}$) exchange energies for the intervalley trion. 

The magnitude of the exchange energies are~\cite{jones2013optical, yu2014dirac}:

\begin{align*}
    J^\textrm{ee}_\textrm{ex} &\approx   g^\textrm{ee} V(\textrm{\textbf{K}}) | \langle u_{\textbf{K}'c\downarrow} | u_{\textbf{K}c\downarrow}  \rangle |^2 \\
     J^\textrm{eh}_\textrm{ex} &\approx   g^\textrm{eh} V(\textrm{\textbf{K}}) | \langle u_{\textbf{K}'v\downarrow} | u_{\textbf{K}c\downarrow}  \rangle |^2 
\end{align*}

where $u$ are the periodic Bloch wave factors at the $K$ and $K'$ points in the Brillouin zone. The indices $c\downarrow$ and $v\downarrow$ denote the spin down conduction band and spin down valence band, respectively, and correspond to the bands shows in Supplementary Fig.~\ref{fig:DFT}b. $g^\textrm{ee}$ and $g^\textrm{eh}$ are the electron-electron and electron-hole contact pair distribution function (PDF), respectively, which correspond to the probability of the finding the particles in the same position. $V(\textrm{\textbf{K}})$ is the unscreened form of the 2d Coulomb interaction.

We calculate the PDFs for 1L-W$_\mathrm{Se}^\mathrm{S}$ by using the quantum Monte Carlo method~\cite{mostaani2017diffusion} and find $g^\textrm{ee}$ to be much smaller than $g^\textrm{eh}$, which is due to the Coulomb repulsion between two electrons. Hence, we neglect $g^\textrm{ee}$ and $J^\textrm{eh}_\textrm{ex}$ dominates over $J^\textrm{ee}_\textrm{ex}$. Using DFT-PBE, we calculate $| \langle u_{\textbf{K}'v\downarrow} | u_{\textbf{K}c\downarrow}  \rangle |^2  = 0.1$ and therefore we calculate the exchange energy splitting of the negative trion to be $J_\textrm{ex} \sim 6$~meV, which agrees well with our observation of an energy difference of 7~meV between X$^-_\textrm{inter}$ and  X$^-_\textrm{intra}$ on device 1 (as discussed in the main text).

\section{Magnetic-field dependent measurements}
\label{sec:magnetic}

We perform a magnetic field dependence of the 1L-W$_\mathrm{Se}^\mathrm{S}$ PL emission, with circularly polarised excitation and detection. Supplementary Figs.~\ref{fig:magnetic}a and b show the PL spectra as a function of magnetic field, under right circularly polarised excitation and for co- and cross-polarised detection, respectively. As we apply the out-of-plane magnetic field, we see the various PL peaks shift in energy. Supplementary Fig.~\ref{fig:magnetic}c shows the co- and cross-polarised (black and red curves, respectively) PL spectra at an applied field of 8~T, with the inset showing the energy splitting for the X$^0$ peak. The energy splitting $\Delta E$  between the opposite circularly polarisations with magnetic field is shown in Supplementary Fig.~\ref{fig:magnetic}d, for the peaks labelled 1, 2 and X$^0$. A linear fitting to the energy splitting gives the g factor as $\Delta E = - g \mu_B B$ for each of the peaks: we find a g factor 4.14(6) for the neutral exciton X$^0$ (as shown in Fig.~4 of the main text), a g factor 5.51(7) for peak 1 and 6.7(1) for peak 2. As discussed in the main text, a g factor $\sim$4 is expected for bright excitons in conventional TMDs. The X$^0$ g factor from PL agrees with that extracted from RC, 4.5(2), and is as expected for the bright neutral exciton. A higher g factor is also seen for lower energy states in  1L-WSe$_2$~\cite{Koperski2018,forste2020exciton, he2020valley} and, similarly to in 1L-WSe$_2$, these peaks likely arise from various excitonic complexes. Full attribution of the different transitions apparent in PL requires further work.

\begin{figure*}
\centering
\includegraphics[width=17.2cm]{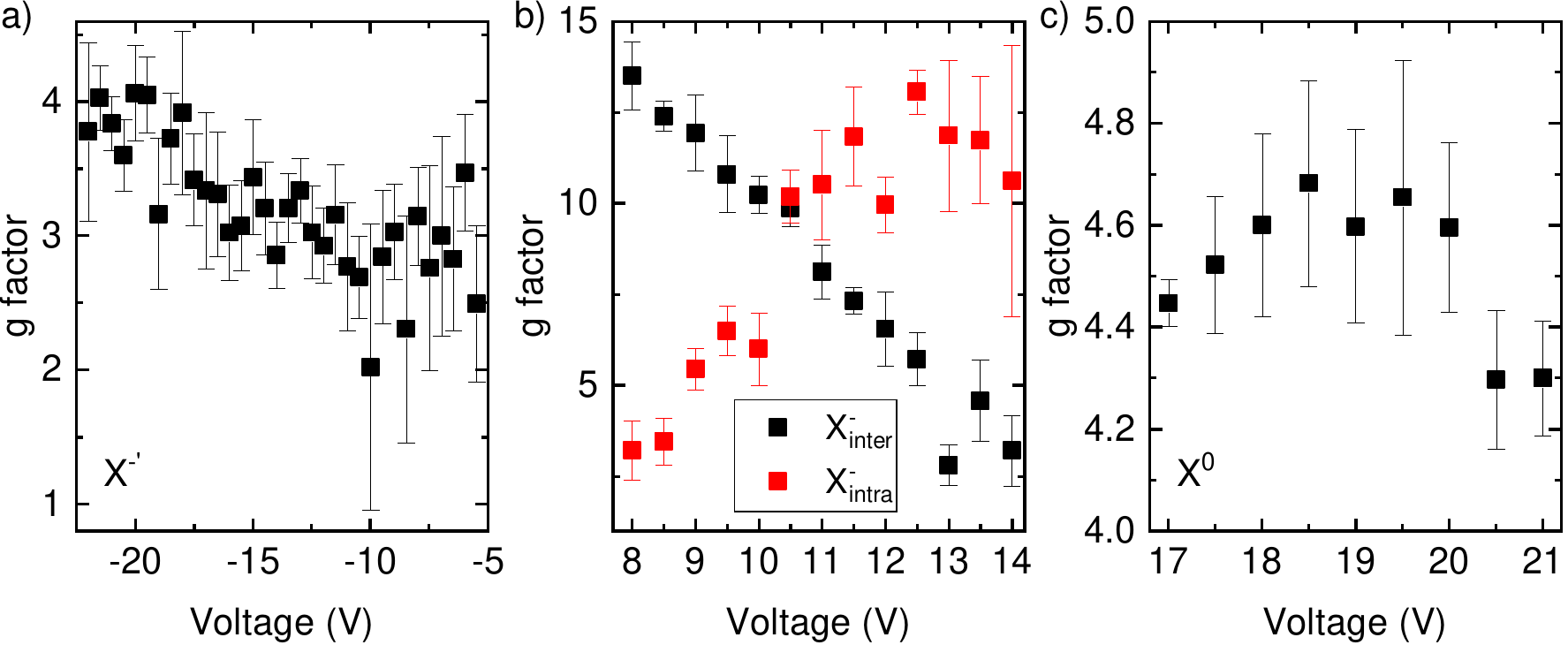}
\caption{{Gate voltage dependence of the g factor of the different excitonic species.}  \textbf{(a)} Voltage dependence of the g factor for the X$^{-\prime}$ peak, between -22 to -5~V. \textbf{(b)} Voltage dependence of the g factor for the X$^-_\textrm{inter}$ (black points) and  X$^-_\textrm{intra}$ (red points) peaks, between 8 to 14~V. \textbf{(c)} Voltage dependence of the g factor for the X$^0$ peak, between 17 to 21~V. All g factors are extracted from a linear fitting to the energy splitting in RC between $\sigma^+$ and $\sigma^-$ detection polarisation with magnetic field. Error bars are given by the fitting error from the linear fitting.}
\label{fig:VdepG}
\end{figure*}

We find that the g factors, as measured in RC, of the various charged complexes shown in Fig.~4 of the main text have a gate-voltage dependence. Supplementary Figure~\ref{fig:VdepG} shows this gate dependence for the X$^{-\prime}$ peak in panel a,  X$^-_\textrm{inter}$ and X$^-_\textrm{intra}$ in panel b (black and red data points, respectively) and X$^0$ in panel c. At each gate voltage, the g factor is extracted from a linear fitting to the magnetic field dependence of the energy splitting of the peak. We see that for  X$^0$ the g factor remains constant at $\sim$4.5 over the entire range of voltages. The g factor of the  X$^{-\prime}$ peak increases as the doping level increases, from $\sim$3 when the peak first appears, towards $\sim$4~V at -20~V. The trion peaks show a larger gate voltage dependence, as discussed in the main text: the inter- and intra-valley trion g factors appear anti-correlated and change from $\sim$3 to $\sim$13. Reported values of the g factors of the negatively charged trions in 1L-WSe$_2$ vary between similar values~\cite{koperski2015single, wang2015magneto, srivastava2015valley, koperski2017optical} and this variation is attributed to the doping dependence of the trion g factor in ref.~\cite{lyons2019valley}. A similar doping dependence of the negatively charged trion in 1L-MoS$_2$ has also been measured and is attributed to many-body interactions with the Fermi sea of electrons~\cite{Klein2021}.

\bibliography{SI.bbl}